\documentclass[twoside,twocolumn,9pt]{article}
\usepackage{extsizes}
\usepackage[super,sort&compress,comma]{natbib}
\usepackage[version=3]{mhchem}
\usepackage[left=1.5cm, right=1.5cm, top=1.785cm, bottom=2.0cm]{geometry}
\usepackage{balance}
\usepackage{times,mathptmx}
\usepackage{sectsty}
\usepackage{graphicx}
\usepackage{lastpage}
\usepackage[format=plain,justification=justified,singlelinecheck=false,font={stretch=1.125,small,sf},labelfont=bf,labelsep=space]{caption}
\usepackage{float}
\usepackage{fancyhdr}
\usepackage{fnpos}
\usepackage[english]{babel}
\addto{\captionsenglish}{%
  
}
\usepackage{array}
\usepackage{droidsans}
\usepackage{charter}
\usepackage[T1]{fontenc}
\usepackage[usenames,dvipsnames]{xcolor}
\usepackage{setspace}
\usepackage[compact]{titlesec}
\usepackage{hyperref}
\usepackage{stackengine}
\usepackage{bm}
\usepackage{siunitx}
\usepackage{physics}
\usepackage{subcaption}
\usepackage{latexsym}
\usepackage{amsmath}
\usepackage{amsfonts}
\usepackage{amssymb}
\usepackage{amsbsy}

\definecolor{cream}{RGB}{222,217,201}

\graphicspath{{./figures/},{./fig/}}

\newcommand{\onlinecite}[1]{\hspace{-1 ex} \nocite{#1}\citenum{#1}}

\begin{document}

\pagestyle{fancy}
\thispagestyle{plain}
\fancypagestyle{plain}{

}

\makeFNbottom
\makeatletter
\renewcommand\LARGE{\@setfontsize\LARGE{15pt}{17}}
\renewcommand\Large{\@setfontsize\Large{12pt}{14}}
\renewcommand\large{\@setfontsize\large{10pt}{12}}
\renewcommand\footnotesize{\@setfontsize\footnotesize{7pt}{10}}
\makeatother

\renewcommand{\thefootnote}{\fnsymbol{footnote}}
\renewcommand\footnoterule{\vspace*{1pt}%
\color{cream}\hrule width 3.5in height 0.4pt \color{black}\vspace*{5pt}}
\setcounter{secnumdepth}{5}

\makeatletter
\renewcommand\@biblabel[1]{#1}
\renewcommand\@makefntext[1]%
{\noindent\makebox[0pt][r]{\@thefnmark\,}#1}
\makeatother
\renewcommand{\figurename}{\small{Fig.}~}
\sectionfont{\sffamily\Large}
\subsectionfont{\normalsize}
\subsubsectionfont{\bf}
\setstretch{1.125} 
\setlength{\skip\footins}{0.8cm}
\setlength{\footnotesep}{0.25cm}
\setlength{\jot}{10pt}
\titlespacing*{\section}{0pt}{4pt}{4pt}
\titlespacing*{\subsection}{0pt}{15pt}{1pt}

\fancyfoot{}
\fancyfoot[LO,RE]{\vspace{-7.1pt}\includegraphics[height=9pt]{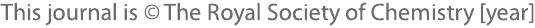}}
\fancyfoot[CO]{\vspace{-7.1pt}\hspace{13.2cm}\includegraphics{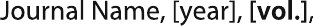}}
\fancyfoot[CE]{\vspace{-7.2pt}\hspace{-14.2cm}\includegraphics{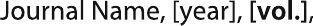}}
\fancyfoot[RO]{\footnotesize{\sffamily{1--\pageref{LastPage} ~\textbar  \hspace{2pt}\thepage}}}
\fancyfoot[LE]{\footnotesize{\sffamily{\thepage~\textbar\hspace{3.45cm} 1--\pageref{LastPage}}}}
\fancyhead{}
\renewcommand{\headrulewidth}{0pt}
\renewcommand{\footrulewidth}{0pt}
\setlength{\arrayrulewidth}{1pt}
\setlength{\columnsep}{6.5mm}
\setlength\bibsep{1pt}

\makeatletter
\newlength{\figrulesep}
\setlength{\figrulesep}{0.5\textfloatsep}

\newcommand{\topfigrule}{\vspace*{-1pt}%
\noindent{\color{cream}\rule[-\figrulesep]{\columnwidth}{1.5pt}} }

\newcommand{\botfigrule}{\vspace*{-2pt}%
\noindent{\color{cream}\rule[\figrulesep]{\columnwidth}{1.5pt}} }

\newcommand{\dblfigrule}{\vspace*{-1pt}%
\noindent{\color{cream}\rule[-\figrulesep]{\textwidth}{1.5pt}} }

\makeatother


\twocolumn[
  \begin{@twocolumnfalse}
\vspace{3cm}
\sffamily
\begin{tabular}{m{4.5cm} p{13.5cm} }

\includegraphics{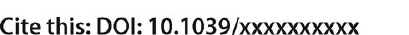} & \noindent\LARGE{\textbf{Wall entrapment of  peritrichous  bacteria: A mesoscale hydrodynamics simulation study$^\dag$}} \\
\vspace{0.3cm} & \vspace{0.3cm} \\

 & \noindent\large{S. Mahdiyeh Mousavi, Gerhard Gompper, and Roland G. Winkler} \\

\includegraphics{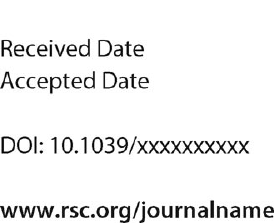} & \noindent\normalsize{%
Microswimmers such as {\em E. Coli} bacteria accumulate and exhibit an intriguing  dynamics near walls, governed by hydrodynamic and steric interactions. Insight into the underlying mechanisms and predominant interactions demand a detailed  characterization of the entrapment  process. We employ a mesoscale hydrodynamics simulation approach to study entrapment of a {\em E. coli}-type cell at a no-slip wall. The cell is modeled  by a spherocylindrical body with several explicit helical flagella. Three stages of the entrapment process can be distinguished: the approaching regime, where a cell swims toward the wall on a nearly straight trajectory; a scattering regime, where the cell touches the wall, with an reorientation of the cell by a torque originating from steric interactions; and a surface-swimming regime. Our simulations show that steric interactions may dominate the entrapment process, yet, hydrodynamic interactions slow down the adsorption dynamics close to the boundary and imply a circular motion on the wall. The locomotion of the cell is characterized by a strong wobbling dynamics, with cells preferentially pointing toward the wall.
}
\end{tabular}

 \end{@twocolumnfalse} \vspace{0.6cm}

  ]

\renewcommand*\rmdefault{bch}\normalfont\upshape
\rmfamily
\section*{}
\vspace{-1cm}


\footnotetext{
Theoretical Soft Matter and Biophysics, Institute of Complex Systems and Institute for Advanced Simulation, Forschungszentrum J{\"u}lich and JARA, D-52425 J{\"u}lich, Germany}




\section{Introduction}

Surfaces  and walls play an essential role in the life cycle of bacteria, because in the wild, bacteria are only rarely isolated and free-swimming, but are primarily associated with surfaces.\cite{cost:95,laug:16,hart:19} In fact, bacteria typically spend most of their life time in a biofilm, rather than as planktonic cell in the bulk fluid, yet biofilm formation is initiated by an initial contact of a planktonic cell with a surface.\cite{flem:16,koo:17} Bacteria approaching a wall experience surface-specific interactions, such as hydrodynamic forces, adhesive forces, steric interactions, etc., which govern the adsorption process and their surface dynamics. The importance of the various interactions for bacteria entrapment  has been addressed experimentally, theoretically, and by simulations.
Studies on wall entrapment of microorganisms, such as non-tumbling \textit{E.~coli} \cite{PhysRevLett.101.038102} and bull spermatozoa \cite{Rothschild1963},  reveal an enhanced concentration at a surface.  This near-wall accumulation of cells can be explained by two distinct  mechanisms.\cite{spagnolie_lauga_2012,elge:16,elge:15} On the one hand, hydrodynamic interactions between the microswimmer and the wall lead to an attractive interaction and, for {\em E.coli}-type bacteria, a torque trying to align the microswimmer with the surface.\cite{PhysRevLett.114.258104, PhysRevLett.101.038102}
On the other hand, neglecting explicit hydrodynamic interactions,  the cell body of a flagellated microswimmer, e.g., {\em E. coli}, moving along a no-slip wall at constant height experiences viscous drag  and a  torque rotates the body.\cite{Happel1983,PhysRevLett.114.258104,das:19.1} At the same time, the  flagellar bundle is exposed to this  torque, which rotates it  away from the wall, and a second torque originated from the  coupling to the cell translational motion. Since the overall system is torque free, a finite inclination angle of the flagellum is obtained with the cell swimming toward the wall.\cite{BiophysJLauga,PhysRevLett.114.258104,laug:16} Finally, in addition to the hydrodynamic mechanisms for the near wall accumulation,  steric interactions with a wall and cell rotational Brownian motion alone can also produce  wall accumulation at finite microswimmer density.\cite{PhysRevLett.103.078101, PhysRevE.84.041932,elge:09}

Despite considerable efforts, the  process of microswimmer entrapment  at walls is by no means satisfactorily described by  explicit modeling so far. Various numerical studies, using the boundary-element method and representing a bacterium by a rigid spheroidal body and the flagella bundle by an attached aligned thin helical cylinder, predict stable configurations of cells swimming at a planar wall.\cite{Shum1725, pimponi_chinappi_gualtieri_casciola_2016}  Cells are found to maintain a stable height above a surface with an inclination angle, where the bacterium's body points away from the wall. Moreover, recent studies on bacteria-like polar swimmer, consisting of a spheroidal body and an active propelling rod, predict a critical swimmer size for entrapment.\cite{spagnolie_lauga_2012} Regardless of the cell body shape, organisms with sufficiently long flagella---about twice the cell-body diameter---are expected to exhibit a positive inclination angle with the head pointing toward the boundary, whereas for short enough flagella, they are orientated away from the wall. This effect can be traced back to a faster increase of the Stokeslet dipole and quadrupole strengths with increasing rod length compared to the source dipole strength.\cite{spagnolie_lauga_2012} Mesoscale hydrodynamics simulations of a mechano-elastic {\em E. coli}-type model with several explicit flagella, including thermal noise and steric interactions, also yield stable, wall-parallel trajectories with a distance of approximately $250 nm$, about half of a body radius,  between the wall and the cell-body  surface.\cite{HuSciRep2015} Similar simulations of a  more complex, swarmer-type bacteria model, \cite{eise:16.1} with an elongated body and a large number of randomly anchored flagella, exhibit stable entrapped trajectories with cells preferentially oriented toward the wall. The elongated nature of the cell body leads to a large (average) angle between the cell body and the inclined bundle. The interplay of near-field hydrodynamic wall interactions and steric repulsion of cell body and bundle determines then the orientation toward the wall.
A recently developed experimental setup enables the full three-dimensional characterization of the entrapment dynamics of smoothly swimming {\em E. coli} bacteria.\cite{PhysRevXDiLeo,bian:19} Such studies clearly reveal a significant angle between the cell body and the surface, with a rather broad angle distribution, but cells pointing toward  the wall. Hence, there is a disparity of available results and lack of a clear understanding of the entrapment mechanism at walls. The insight into the corresponding cell-level  processes  is not only fundamental for biological systems, e.g., biofilm initiation, but  should also be essential for the rational design of  biomimetic microrobots. \cite{pala:18,gomp:20}

In this article, we perform mesoscale-hydrodynamics simulations of a mechano-elastic {\em E. coli}-type model with several flagella to shed light onto the entrapment of bacteria at a no-slip surface. Our bacterium model closely resembles the geometry, flagellar elastic properties, and rotary motor torque of {\em E. coli}  with  multiple flagella.\cite{hu:15.1,eise:16.1} A hybrid simulation approach is adopted, where molecular dynamics simulations for the bacterium  are combined with  the multiparticle collision dynamics (MPC) method for the embedding fluid.\cite{ripo:04} MPC is a particle-based simulation approach taking into account hydrodynamic interactions and thermal fluctuations.\cite{male:99,kapr:08,gomp:09} It is a valuable and adequate simulation method to  study nonequilibrium, active systems.\cite{goet:10,kapr:08,elge:09,earl:07,elge:10,
babu:12,elge:13,thee:13,yang:14,zoet:14,elge:15,hu:15.1,HuSciRep2015,eise:16.1,qi:20}  In particular, MPC has successfully been used to characterize synchronization of flagella beating between nearby swimming sperm, \cite{yang:08} bundling of helical flagella of bacteria,\cite{reig:12,reig:13} swimming of bacteria near walls, \cite{HuSciRep2015,eise:16.1}, and studies on clustering of squirmers.\cite{thee:18}

In our simulations, the cell entrapment process can be classified in three regimes, the approaching regime, where cells swim essentially straight toward the surface, the scattering regime, where they reorient, and the surface swimming regime, where they swim along the wall on circular trajectories. We confirm that  hydrodynamic interactions reduce the swimming velocity while they approach the surface, however, it is rather due to lubrication than hydrodynamic interactions by the force-dipole flow field. Similarly, reorientation is caused by steric cell-surface interactions and/or hydrodynamics.\cite{dres:11,bian:19} In contrast to the various theoretical predictions, our cells preferentially point toward the surface during stable near wall locomotion. Since we consider  randomly anchored flagella on the cell surface, we find a wide spectrum of wobbling motions, where the cell body's orientation varies periodically.\cite{hyon:12} Interestingly, the bundle orientation and overall swim direction is little affected by the strong orientational fluctuations of the cell body. Evidently, the flagella bundle rotation determines the overall cell swimming behavior and the cell body responds  to the this rotation. 

The manuscript is organized as follows. In Sec.~\ref{sec:model}, the cell model and simulation approach are described. Section~\ref{sec:results} presents the results for the various stages of the scattering process, and Sec.~\ref{sec:summary} summarizes our findings.

\section{Simulation method} \label{sec:model}
\subsection{\textit{E}.~coli model}

\begin{figure}[t]
	\centering
	\includegraphics[width=\columnwidth]{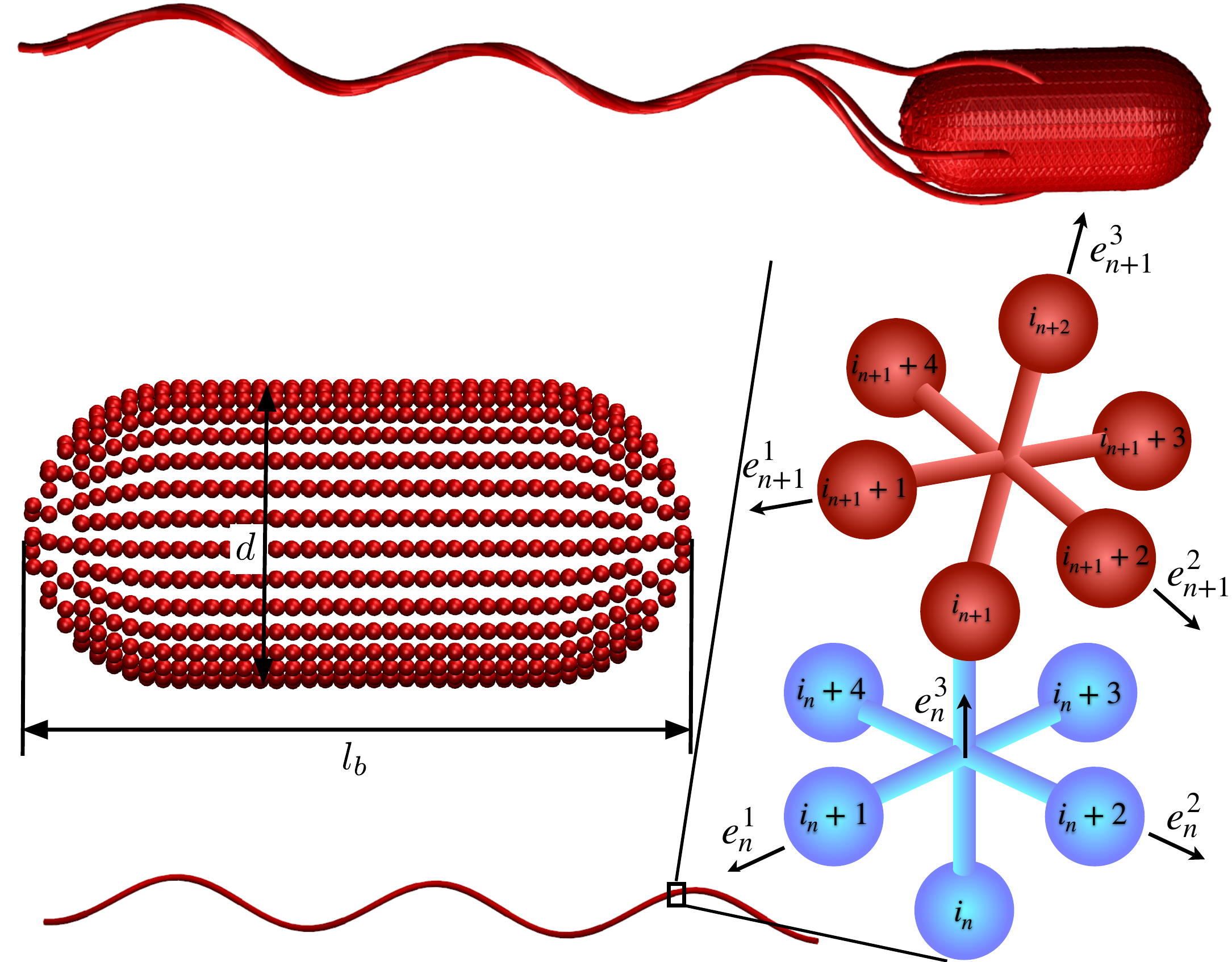}
	\caption{Model of an {\em E. coli} cell with a  spherocylindrical cell body of diameter $d = 0.9 \mu m$ and length $l_b = 2.5 \mu m$. It is composed of circular sections of particles, which are connected by the bond potential of Eq.~\eqref{eq:bondPoten}. The flagellum, a three-turn left-handed helix of radius $R = 0.2 \mu m$, pitch $L = 2.2 \mu m$ and contour length $L_c = 7.6 \mu m$ consists of 76 consecutive octahedral-like  segments.}
	\label{fg:cell}
\end{figure}

We employ a mechano-elastic bacterium model.\cite{hu:15.1,HuSciRep2015,eise:16.1}. The cell is composed of a spherocylindrical body and flagella modelled by semiflexible helical filaments (cf. Fig.~\ref{fg:cell}).
The whole cell is constructed by points of mass $M$. For the body, the points are arranged on the circumference and in the center of circles of diameter $d=9a$, $a$ is the length unit related to the MPC fluid described in Sec.~\ref{sec:MPC}, with a spacing of $0.5a$ (along the cylinder center line) and on smaller circles toward the poles. The cylindrical part consists of $22$ and the spherical cap of nine such circles corresponding to the body length $l_b=20 a$.  In order to preserve the stability of the body, nearest and next-nearest-neighboring particles are connected by a harmonic potential of the form
\begin{equation}
	U_b = \frac{1}{2}K_b(r-r_e)^2 ,
	\label{eq:bondPoten}
\end{equation}
where $r$ and $r_e$ are the distance between a particle pair and its preferred value, respectively, and $K_b$ is the bond strength. Each bacterium is equipped with 5 randomly anchored flagella, which are described by the helical wormlike chain model,\cite{HWCh1997, Vogel2010} with an adaptation suitable for the combination with MPC.\cite{hu:15.1} A flagellum consists of $N = 76$ segments of $6$ particles arranged in an octahedral shape, with $12$ bonds along the edges of length $a/\sqrt{2}$ and $3$ along the diagonals of length $a$. This construction allows for the intrinsic twist of the flagellum and its coupling to MPC particles capturing  torsional fluctuations.

Introducing the  orthogonal  bond vectors  $\bm{b}_n^1 = \bm{r}_{i_n+1} - \bm{r}_{i_n+3}$, $\bm{b}_n^2 = \bm{r}_{i_n+2} - \bm{r}_{i_n+4}$, and $\bm{b}_n^3 = \bm{r}_{i_n+1} - \bm{r}_{i_n}$ for the bonds along the contour of the flagellum, we can define orthonormal triads $\{\bm{e}_n^1, \bm{e}_n^2,\bm{e}_n^3\}$, $n=1,\ldots,N$, where $\bm{e}_n^{\alpha} = \bm{b}_n^{\alpha}/|\bm{b}_n^{\alpha}|$,  $\alpha \in \{1,2,3\}$ (cf. Fig.~\ref{fg:cell}). The local elastic deformation of a flagellum proceeds  in two steps: (i) the rotation of $\{\bm{e}_n^1, \bm{e}_n^2,\bm{e}_n^3\}$ around $\bm{e}_n^3$ by a twist angle $\varphi_n$ and (ii) the rotation of the twisted triad $\{\tilde{\bm{e}}_n^1, \tilde{\bm{e}}_n^2,\tilde{\bm{e}}_n^3\}$  by a bending angle $\vartheta_n$ around the normal $\bm{n}_n = (\bm{e}_n^3\times \bm{e}_{n+1}^3)/|\bm{e}_n^3\times \bm{e}_{n+1}^3|$ to the plane defined by the contour bonds $\bm{b}_n^3$ and $\bm{b}_{n+1}^3$. The elastic deformation energy is then
\begin{equation}
	U_{el}=\frac{1}{2}\sum_{\alpha=1}^{3}K_{el}^{\alpha}\sum_{n=1}^{N-1}(\Omega_n^{\alpha} - \Omega_e^{\alpha})^2,
	\label{eq:elasPot}
\end{equation}
where $K_{el}^1=K_{el}^2$ is the bending modulus, $K_{el}^3$ the twist modulus, and $\bm{\Omega}_n = \Omega_n^1\bm{e}_n^1 +  \Omega_n^2\bm{e}_n^2 + \Omega_n^3\bm{e}_n^3 \equiv \vartheta_n\bm{n}_n + \varphi_n \bm{e}_n^3$ the strain vector. The parameters $\Omega_e^{\alpha}$ define the equilibrium geometry of the model flagellum and are chosen to recover the shape of an \textit{E}.~coli flagellum in the normal state. \cite{Darnton1756}

A flagellum is attached to the body by randomly choosing a body particle as its first contour particle. The rotation of the flagellum is induced by a motor torque $\bm{T}$ decomposed into a force couple $\bm{F}$ and $-\bm{F}$ acting on particles $i_1+2$ and $i_1+4$ ($\bm{T} = \bm{b}_1^2 \times \bm{F}$ with $\bm{F}$ parallel to $\bm{b}_1^1$), or equivalently $i_1+1$ and $i_1+3$  ($\bm{T} = \bm{b}_1^1 \times \bm{F}$ with $\bm{F}$ parallel to $\bm{b}_1^2$). Moreover, an opposite torque $-\bm{T}$ is exerted on two body particles non-aligned  with the body axis and on different circles in the vicinity of the anchoring point. Hence, the bacterium is force and torque free. A repulsive harmonic potential
\begin{equation}
U_{ex}=\begin{cases}\frac{1}{2}K_{ex}(r-r_{ex})^2 &  r < r_{ex}
\\ 0 & \text{otherwise}\end{cases}
\label{eq:repulsive_harmonic}
\end{equation}
is used to prevent flagella crossing and their penetration into the cell body. Here, $r$ is the closest distance between contour bond segments of different flagella and the distance to the body-center particles. We set $r_{ex} = 0.25a$ and $r_{ex} = (d+a)/2$ for the flagellum-flagellum and flagellum-body interactions, respectively.

The forces resulting from the potentials \eqref{eq:bondPoten}--\eqref{eq:repulsive_harmonic} and the forces induced by the torques $\bm{T}$ and $-\bm{T}$ determine the dynamics of the bacterium, which is described by  Newton's equation of motion. The latter are solved by the velocity Verlet integration scheme.\cite{alle:87}

\subsection{Fluid model: multiparticle collision dynamics}\label{sec:MPC}
The fluid is modeled as a collection of point particles of mass $m$ with position $\bm{r}_i$ and velocity $\bm{v}_i$. The dynamics of the particles proceeds by alternating streaming and collision steps.\cite{male:99,kapr:08,gomp:09} During a streaming step, particles move ballistically over a time interval $\Delta t$, denoted as collision time, and the their positions are updated according to
\begin{equation}
\bm{r}_i(t+\Delta t) = \bm{r}_i (t) + \bm{v}_i (t)\Delta t .
\label{eq:MPCstreaming}
\end{equation}
 In the collision step, all particles are sorted in cubic collision cells of length $a$. Subsequently, the relative velocity of each particle, with respect to the center-of-mass velocity of the considered collision cell, is rotated by a fixed angle $\alpha$ around a randomly oriented axis, hence, their velocities after the collision are \cite{thee:16}
\begin{align} \nonumber
	\bm{v}_i(t + \Delta t) = \bm{v}_{cm}(t) + \mathrm{\bf R}(\alpha)  \big[\bm{v}_i(t) - \bm{v}_{cm}(t) \big] - \bm{r}_{ic} \\
	\times \Big[m \mathrm{\bf I}^{-1} \sum_{j\in cell} \big( \bm{r}_{jc}(t)\times [ \bm{v}_{jc}(t) - \mathrm{\bf R}(\alpha)\bm{v}_{jc}(t) ] \big)  \Big] ,
	\label{eq:MPCcollision}
\end{align}
where  $\mathrm{\bf R}(\alpha)$ is the rotation operator,  $\bm{v}_{cm}$  the center-of-mass velocity, $\bm{r}_{jc} = \bm{r}_j - \bm{r}_{cm}$, $\bm{v}_{jc} = \bm{v}_j - \bm{v}_{cm}$, and $\mathrm{\bf I}$ the moment-of-inertia tensor of the particles in the center-of-mass reference frame of the collision cell of particle $i$. The collision rule \eqref{eq:MPCcollision} conserves both linear and angular momentum in each cell.\cite{nogu:08,thee:14,thee:16}
Discretization in collision cells implies violation of Galilean invariance, which is reestablished by
a random shift of the collision-cell grid after every streaming step.\cite{ihle:03} A constant temperature is maintained by  a collision-cell-based, local Maxwellian thermostat, where the relative velocities of the particles in a collision cell are scaled according to the Maxwell-Boltzmann scaling (MBS) method.\cite{huan:15}

\subsection{Coupling of bacterium and MPC fluid}

The coupling between the MPC fluid and the bacterium is achieved in the MPC collision step by treating the points of the bacterium on equal footing with the MPC particles, i.e., their velocities are also rotated according Eq.~(\ref{eq:MPCcollision}) to ensure momentum exchange between them and the fluid.\cite{ripo:04,muss:05,eise:16.1} Here, the center-of-mass velocity of a collision cell is given by
\begin{align} \label{center-of-mass_cell}
{\bm v}_{cm}=\frac{1}{mN_c+ M N_c^c} \left( \sum_{i =1}^{N_c} m {\bm v}_i + \sum_{j=1}^{N_c^c} M {\bm v}_j^b \right) ,
\end{align}
where, $N_c^c$ is the number of mass points of a bacterium in the considered collision cell. Note that the cell body  is penetrable for fluid particles by this coupling. However, this still provides a no-slip boundary condition on the body surface.\cite{pobl:14}

\subsection{Wall interaction}

The fluid is confined between two walls parallel to the $xz$ plane of the Cartesian reference frame.  No-slip
boundary conditions are implemented by employing the bounce-back rule for MPC particles and by taking into account  wall phantom particles.\cite{lamu:01,gomp:09,huan:15} To avoid direct wall contact,  a cell experiences the repulsive Lennard-Jones potential (wall at $y=0$)
\begin{align}
U_{W} = \left\{
\begin{array}{cc}
4 k_BT \left[ \left( \displaystyle \frac{\sigma}{y-y'} \right)^{12} - \displaystyle \left( \frac{\sigma}{y-y'} \right)^6 + \frac{1}{4} \right]    & y -y' < y_c \\
0  & \text{otherwise}
\end{array} ,
\right.
\end{align}
where $y$ is either the distance between a flagellum contour particle and the wall, or that of a body center-line particles and the wall. Hence,  $y' = d/2$ for cell body and $y' = 0$ for a flagellum particle.

\subsection{Parameters}

We choose $K_{el}^1 = K_{el}^2 = K_{el}^3 = 5 \times 10^{4}k_BT$ within the range of experimentally measured  values of the flagellar filaments.\cite{Darnton1756,hu:15.1,Vogel2010} Here, $k_B$ is the Boltzmann factor and $T$ the temperature.  The magnitude of the torque is set to  $|\bm{T}| = 400k_BT = 1640~pN\,nm$, smaller than the torque measured experimentally ($4500~pN\,nm$)  for a stalled motor.\cite{Berry14433, HuSciRep2015,das:18.2} As a consequence, the bundle rotates with the average  frequency $\omega_{bund} = 3.14 \times 10^{-2} \sqrt{k_BT/ma^2}$, five times faster than the body, comparable to {\em E.~coli} cells.\cite{Darnton1756} The force constant of the bonds and the repulsive harmonic potential are set to $K_b=K_{ex}=10^4k_BT/a^2$. The cut-off distance for the wall interaction is $y_c =\sqrt[6]{2} a$ and the interaction range is $\sigma=a$.

Length and time are measured in units of the collision cell size $a$ and $\tau = a\sqrt{m/k_BT}$, respectively. We choose the collision time $\Delta t = 0.05\tau$ and the average number of fluid particles in a cell $\langle N_c \rangle = 10$. A cubic simulation box of length $200a$ is considered, with two no-slip walls parallel to the $xz$ plane of the Cartesian reference frame and periodic boundary conditions parallel to the walls.  At least  ten independent realizations are considered for every shown parameter set.

\begin{figure}[t]
	\centering
	\includegraphics[width=\columnwidth]{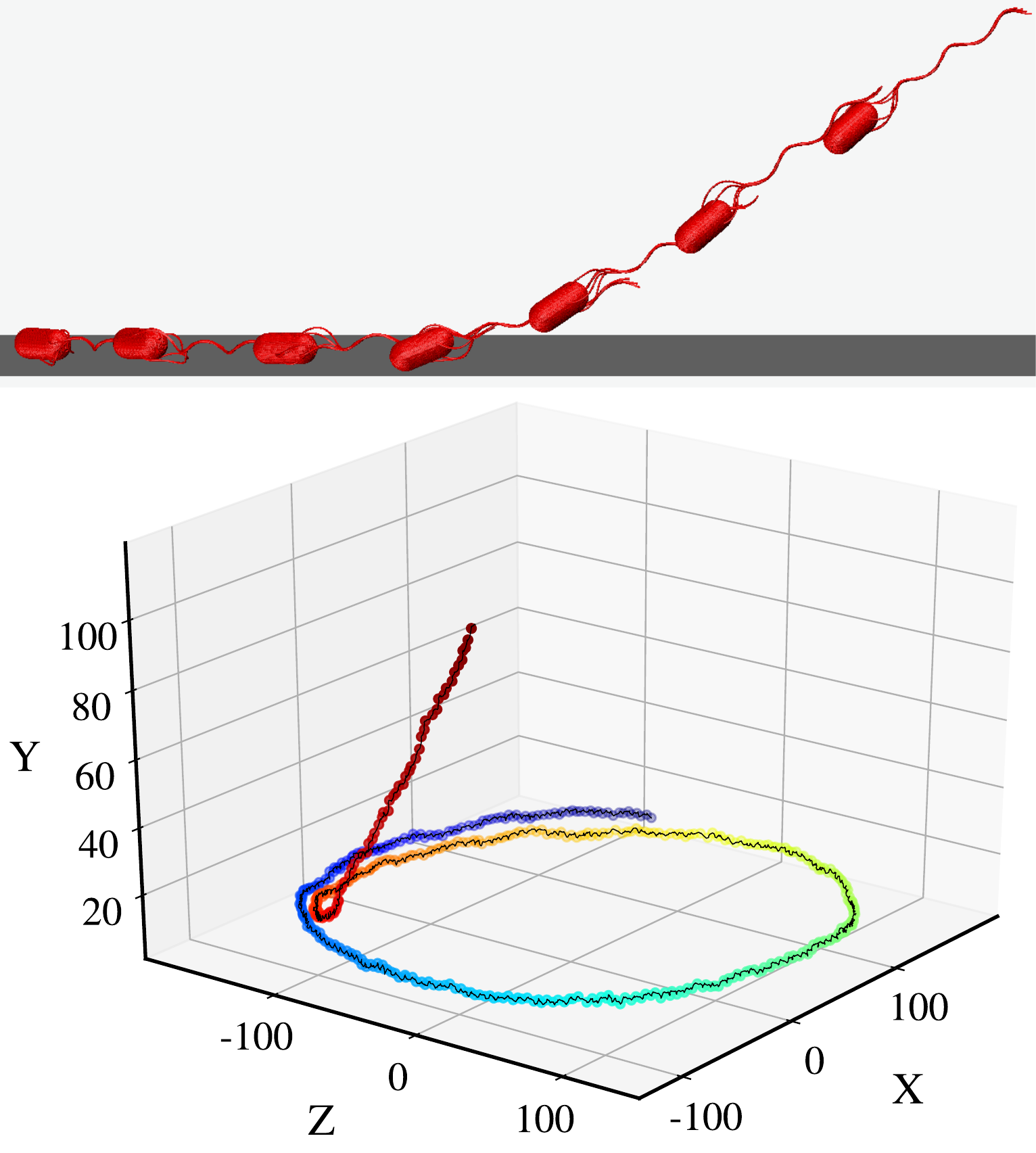}
	\caption{Trajectory of a cell approaching a wall and undergoing a clockwise circular motion on the surface. The starting  angle is $\theta_s=45^{\circ}$. (See supplementary files, movie M1)}
	\label{fig:traj}
\end{figure}

\section{Results} \label{sec:results}

Initially, swimming cells are created by independent bulk simulations, with randomly anchored, radially outward pointing flagella. By applying independent torques $\bm T$ to rotate flagella, a bundle is formed and a cell starts to swim uni-directionally. Subsequently, the cell is placed in the center between the two walls and rotated such that its swimming direction forms an angle $\theta_s$ with the wall. The cell orientation of the subsequent trajectory is characterized by the inclination angle $\theta_i(t)$ between the main axis of the inertia tensor of the cell (body and  flagella)---in fact, it closely agrees with the orientation of the flagellar bundle as well as the swimming direction---and its projection onto the horizontal wall (cf. Fig.~\ref{fig:traj}) ($\theta_s=\theta_i(0)$). 

As is well established, motile bacteria display helical swimming paths, \cite{liu:14} since the flagellar bundle is typically not aligned with the cell body, but can be strongly inclined.\cite{eise:16.1,darn:07.1,turn:10} The inclination causes a wobbling motion of the cell body, i.e., it precesses around the swimming direction. \cite{darn:07.1,Patteson_2015, hyon_marcos_powers_stocker_fu_2012,PhysRevXDiLeo,bian:19} Wobbling (wiggling) depends on different cell parameters such as the orientation and position of the flagellar bundle relative to the cell body, or the viscoelastic properties of the surrounding fluid.\cite{Patteson_2015, hyon_marcos_powers_stocker_fu_2012} We expect an effect of wobbling on the cell-surface scattering process.

\begin{figure*}[t]
		\includegraphics[width=\textwidth]{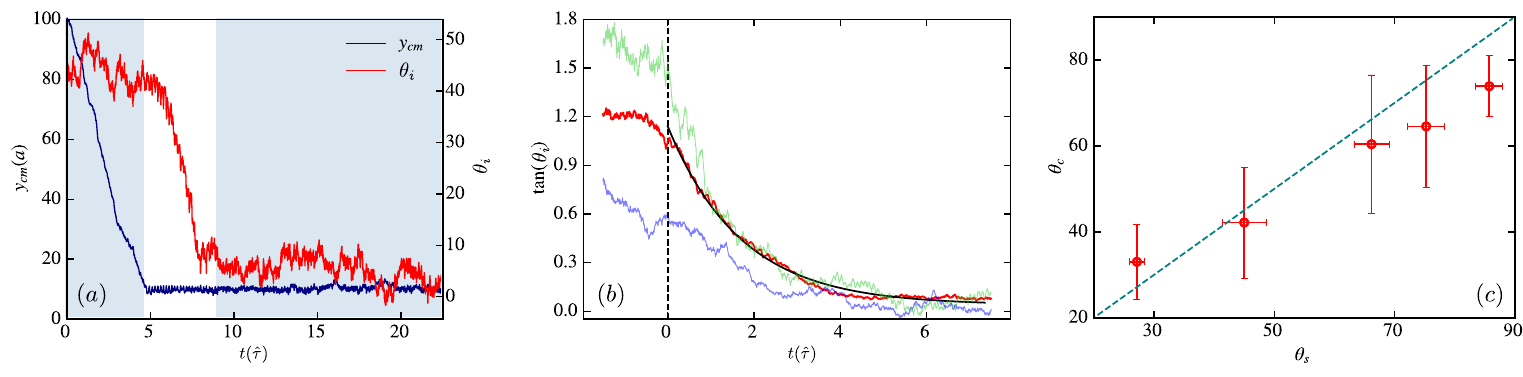}
	\caption{(a) Time evolution of the center-of-mass position of the cell body above the wall, $y_{cm}$, and the inclination angle, $\theta_i$, during a wall-entrapment event. (b) Time evolution of the inclination angle during the reorientation stage for a strongly (green) and weakly (blue) wobbling cell for the initial angle $45^{\circ}$. The red curve is the average value over 14 realizations. The  dashed line corresponds to the time of collision, $t_c$, with the wall, and the black solid line is a fit to Eq.~\eqref{eq:theta_orient}.  (c) Inclination angle at the first encounter with the wall as function of its initial value. The bullets represent average value over all realizations and error bars are the standard deviation. The dashed line indicates $\theta_i = \theta_s$.}
	\label{fig:entrapment}
\end{figure*}

Figure~\ref{fig:traj} depicts a cell approaching a wall and its clockwise swimming at the wall. The time dependence of the center-of-mass height above the wall is displayed in Fig.~\ref{fig:entrapment}. Initially, a cell swims more or less straight toward the wall until it interacts with the wall. Subsequently, the cell moves along the wall at a nearly constant height in a circular manner.

Three regimes can be identified in Fig.~\ref{fig:entrapment}(a): (i) The approaching regime $t/\hat \tau \lesssim  5 $, where the inclination angle is fluctuating, but the cell swims essentially straight,  (ii) the reorientation regime $ 5 \lesssim t/\hat \tau \lesssim 9$, and (iii) the surface swimming regime $t/\hat \tau  \gtrsim 9 $. Here, $\hat \tau = l_b/\bar v = 0.67 \times 10^4 \tau$ is the time required for a cell with the average velocity $\bar v = 3\times10^{-3} \sqrt{k_BT/m} \approx 10 \mu m/s$ to swim over its body length $l_b$.

\subsection{Wall approach} \label{sec:approach}

\begin{figure}[t]
		\includegraphics[width=\columnwidth]{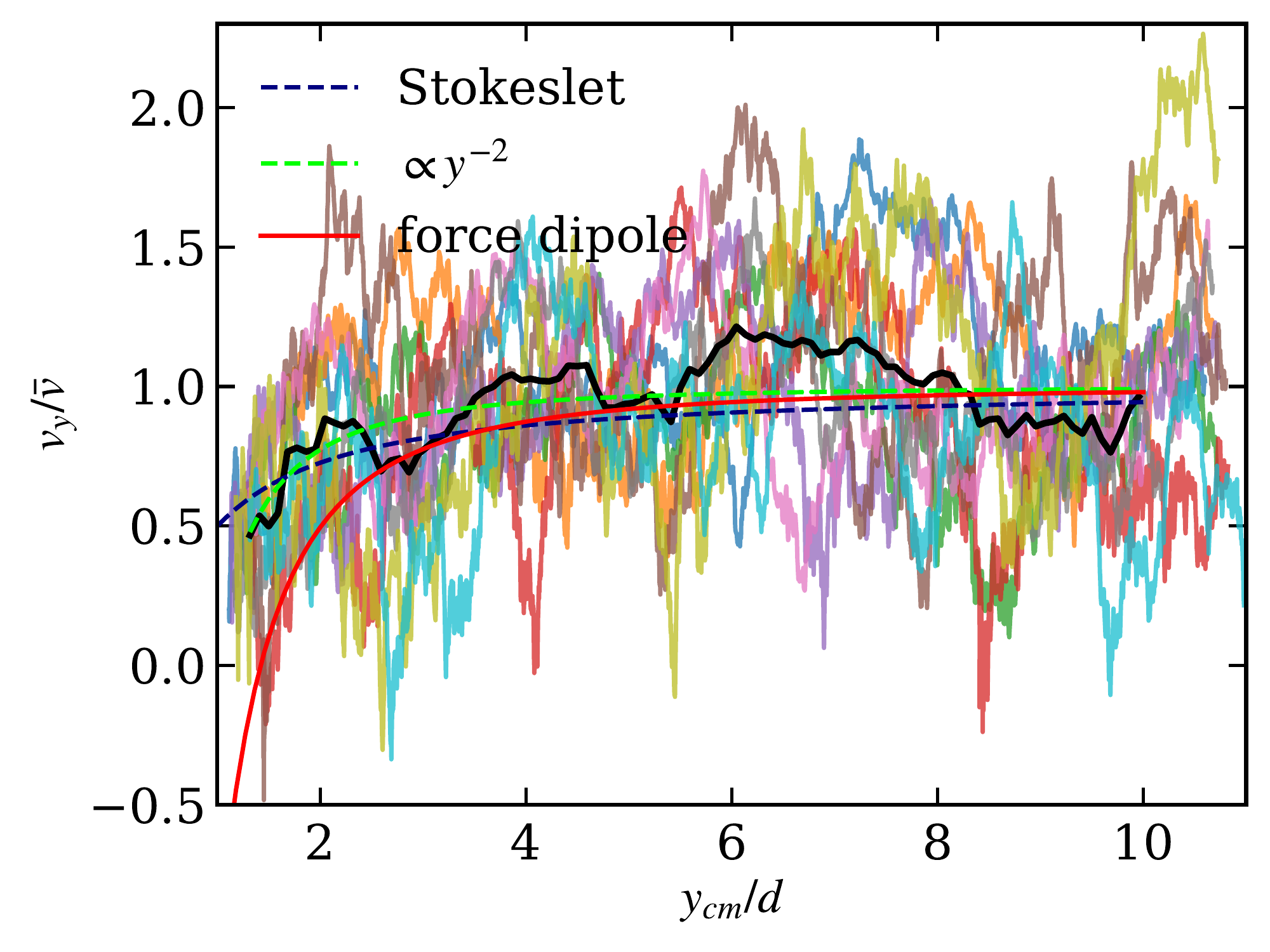}
\caption{Velocity normal to the wall as a function of the body center-of-mass position above the wall for various realization and their average (black line); the initial angle is $\theta_s \approx 45^{\circ}$. The velocities are normalized by the swimming velocity, $\bar v$, far from the wall. The purple dashed line is obtained with  the Stokeslet  velocity of  Eq.~\eqref{eq:velocity_stokes} The red dashed line is calculated with the  force-dipole contribution \eqref{eq:force_dipole}, and  the green dashed line is a fit of the velocity profile including Eq.~\eqref{eq:force_dipole}. }
	\label{fig:velocity}
\end{figure}

Our simulations yield an effect of the wall on the cell swimming velocity in its vicinity.   As shown in Fig.~\ref{fig:velocity}, we find a slowdown of the velocity normal to the wall already for cell-wall separations larger than a body length. 

An important aspect in cell entrapment  is  the role of hydrodynamic interactions, a fact, which   has been addressed in various experiments.\cite{dres:11,bian:19} Experimental studies lead to the conclusion that hydrodynamic interactions are of minor importance.\cite{dres:11}  However, other studies  show a reduction of the velocity in the direction normal to the surface of an approaching bacteria when closer than approximately a cell length,\cite{PhysRevXDiLeo,bian:19,frym:95}  which is attributed to an increased friction due to hydrodynamic interactions.\cite{bian:19,rami:93}
The far-field approximation for an approaching (externally pulled) sphere yields the normal velocity at a no-slip wall 
\begin{align} \label{eq:velocity_stokes}
v_y = \bar v \left(1 - \frac{9}{16}\frac{d}{y_{cm}} +\frac{1}{16} \left(\frac{d}{y_{cm}}\right)^3 \right) ,
\end{align}
where $\bar v$ denotes the velocity far from the boundary (bulk) and $y_{cm}$ the location of the sphere's center.\cite{bian:19,rami:93,kim:91}  Our simulation results are in  agreement with Eq.~\eqref{eq:velocity_stokes} within the accuracy of the simulation, cf. Fig.~\ref{eq:velocity_stokes}. 

However, Eq.~\eqref{eq:velocity_stokes} corresponds to the velocity of an approaching Stokeslet, whereas the near-flow field of an {\em E. coli} is very different from the Stokeslet flow field.\cite{hu:15.1} A more suitable approximation would be the normal velocity  emerging from the  interaction of a hydrodynamic dipole with a no-slip wall, namely 
\begin{align} \label{eq:force_dipole}
v_y^d = - \frac{3p}{64 \pi\eta y^2}\left(1 - 3 \sin^2 \theta_i \right) ,
\end{align}
where $y$ is the height of the center of the considered force dipole above the wall, $p$ the dipole strength, where $p>0$ for {\em E. coli}, $\eta$ the fluid viscosity, and $\theta_i$ the inclination angle.\cite{PhysRevLett.101.038102,spagnolie_lauga_2012,wink:18}  The actual  velocity in then $v_y = \bar v +v_y^d$. Normalization with respect to the average swimming velocity $\bar v$ and the diameter of the cell body, yields the relation 
\begin{align} \label{eq:s_force_dipole}
\frac{v_y}{\bar v} =1 - u_0 \left(\frac{d}{y}\right)^2 .
\end{align}
Using the MPC fluid parameters and the dipole strength $p=2.2 pN \mu m$, obtained in Ref.~\onlinecite{hu:15.1} for swimming {\em E. coli}, we find $u_0\approx 2$. As shown in Fig.~\ref{fig:velocity}, the velocity contribution of  Eq.~\eqref{eq:s_force_dipole} with $u_0=2$ drops somewhat faster for $y_{cm} \to 0$ than the simulation data. However, the data are very well reproduced by fitting Eq.~\eqref{eq:s_force_dipole}, which yields $u_0 = 0.9$. Considering the uncertainties in the determination of the force dipole, specifically in the vicinity of the wall, $u_0=2$ is reasonably close to the fitted value, which suggest that the slow-down might be explained by  hydrodynamic  interactions of the self-propelled cell with the wall. Unfortunately,  our data are not precise enough to rule out one of the interactions mechanisms (Eqs.~\eqref{eq:velocity_stokes}, \eqref{eq:s_force_dipole}). 


\subsection{Impact and scattering}

The time dependence of the inclination angle during cell reorientation is displayed in Fig.~\ref{fig:entrapment}(b) for two  realizations of random flagella anchoring, which emphasizes the dependence on the extent of wobbling, as well as the average over several realizations.
Two mechanisms contribute to cell reorientation: (i) hydrodynamic interactions with the wall for a force and torque free swimmer and (ii) steric interactions as soon as the swimmer experiences mechanical wall forces. 

 Hydrodynamic interactions yield the rotation frequency
\begin{align} \label{eq:omega_hydro}
	\Omega_z(\theta_i , y) = -\frac{3p\cos\theta_i \sin\theta_i}{64\pi \eta y^3}\left[1 + \frac{(\gamma^2-1)}{2(\gamma^2+1)}(1 + \sin^2\theta_i) \right] ,
\end{align}
where $\gamma$ is the aspect ratio of the cell $(\gamma > 1)$.\cite{PhysRevLett.101.038102,spagnolie_lauga_2012,wink:18} The influence of the hydrodynamic reorientation on cells can be deduced from Figs.~\ref{fig:entrapment}, displaying the time dependence of the inclination angle (Fig.~\ref{fig:entrapment}(b)) and the inclination angle at surface contact, $\theta_c$, as function of the initial angle $\theta_s$ (Fig.~\ref{fig:entrapment}(c)). Figures~\ref{fig:entrapment}(a), (b) suggest a certain reorientation before a cell encounters the wall, however, by a few degrees only. Similarly, Fig.~\ref{fig:entrapment}(c) shows small differences between $\theta_s$ and $\theta_c$  at low $\theta_s$ values,  but $\theta_c$ drops below $\theta_s$ at large $\theta_s$ as in experiments.\cite{bian:17}  

A quantitative comparison of Eq.~\eqref{eq:omega_hydro} with the simulation results  is hampered by the strong dependence of $\Omega_z$ on the height $y$ above the wall. To account for cell body and flagellar bundle, we use the body length for $y=l_B$, which is approximately equal to the hydrodynamic radius of the cell.\cite{dres:11}  
With the value $p=2.2 pN \mu m$ for the force dipole and $\gamma= 2.2$ for the aspect ratio of the cell body, Eq.~\eqref{eq:omega_hydro} yields $\Omega_z \hat \tau \approx - 0.8 rad \approx - 45^{\circ}$ for $\theta_i =45^{\circ}$. This value is approximately $3$ times larger than the change in the angle obtained in simulations, $15^{\circ}$, over this time scale,  but it is on the right order of magnitude.  Hence, hydrodynamic interactions between the swimmer flow field and the wall might play a role in the reorientation  dynamics during surface scattering, in contrast to conclusions based on experiments.\cite{dres:11,bian:19}.

Steric interactions yield reorientation of the cell after it touched a wall. While interacting with the wall, the cell experiences a repulsive normal force,  which is $F_s^{\perp} = - \gamma_T \bar v \sin \theta_i$, i.e., the cell is  no longer force free;  $\gamma_T$ is the translational friction coefficient.  With the assumption  that the propulsion force acts on the center of mass of the cell located at a distance $l_{cm}$ separated from the cell-surface contact point, $F_s^{\perp}$ implies the torque $M= - \gamma_T \bar v l_{cm} \sin \theta_i \cos \theta_i$. Then, the equation of motion $\gamma_R \dot \theta_i =  M$, $\gamma_R$ is the  rotational friction coefficient, yields
\begin{align}
\frac{d}{dt} \tan \theta_i = -\gamma_i \tan \theta_i ,
\end{align}
and, thus, the time dependence
\begin{align} \label{eq:theta_orient}
\tan \theta_i (t) = \tan \theta_i (0) e^{-\gamma_i t} ,
\end{align}
where $\gamma_i=\gamma_T \bar v l_{cm}/\gamma_R$. Figure~\ref{fig:entrapment}(b) shows a  fit of Eq.~\eqref{eq:theta_orient} to the average of the  simulation data. Evidently, the reorientation dynamics is well described by Eq.~\eqref{eq:theta_orient} over the considered time interval for $\gamma_i= 0.6/\hat \tau \approx 3/s$. (For the translation of simulation units to physical units, cf. Ref.~\onlinecite{hu:15.1}.) Using the above theoretical expression for $\gamma_i$, we find $\gamma_i \approx 7.5/s$. Within  the uncertainties in the parameters, this value agrees reasonably well with the fit. Moreover, our values are in very good agreement with those obtained in the experiments, \cite{bian:19} where a fit yields $\gamma_i \equiv \kappa' \approx 4.9/s$, and a theoretical estimation $\kappa' \approx 6.3/s$. Thus,  the angular momentum, appearing as a consequence of steric interaction as soon as a cell touches a wall, can be responsible for cell reorientation, in agreement with previous studies.\cite{dres:11,bian:19} However, no definite conclusion is possible based in the simulation data, because of the wide uncertainty of parameters entering the rather generic  theoretically expressions. For a more precise estimate, a more detailed theoretical approach is needed.

\begin{figure*}[t]
\includegraphics[width=\textwidth]{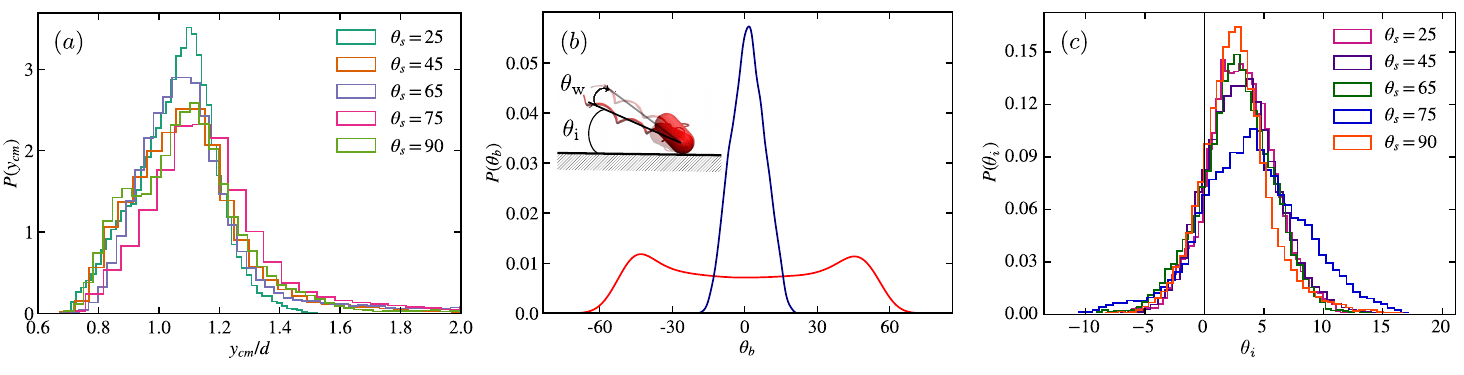}
\caption{(a) Distribution function of the body center-of-mass distance from the wall for different initial angles. (b) Distribution of inclination angle for cells with a large (red) and low wobbling (blue) angle. Inset: Definition of inclination and wobbling angle of a cell swimming close to a wall. (c) Distribution function of the inclination angle for cells swimming at surfaces for the indicated starting angles.}
\label{fig:pitch_dist}
\end{figure*}

\begin{figure}[t]
\includegraphics[width=\columnwidth]{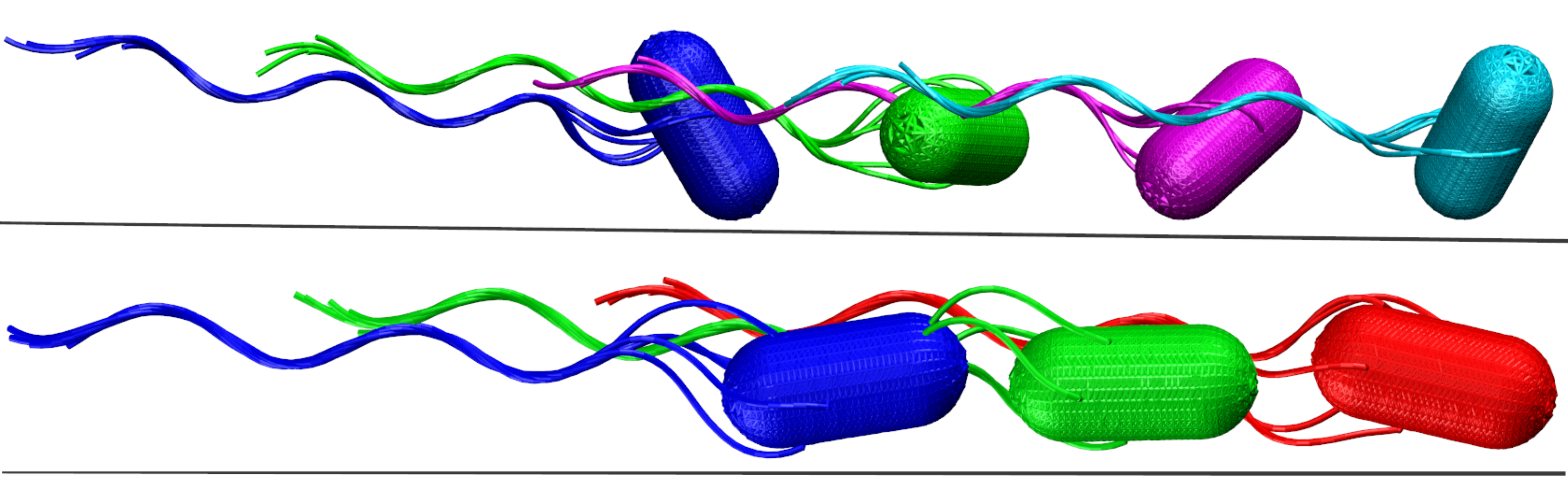} \\
\caption{Snapshots of cells swimming at a wall (black line) for two realizations of flagella arrangements. (Top) The flagella are mainly localized along the equator of the cell body. This leads to a very strong wobbling motion. (Bottom) Example of a cell with flagella  preferentially anchored at the rear part of the cell body, which implies weak wobbling. (See supplementary files, movie M2 and M3.)}
\label{fig:snapshot_cell_orient}
\end{figure}

\subsection{Surface swimming}

During stationary surface migration, the cells progress with their body center-of-mass at nearly constant height. The center-of-mass distribution function, Fig.~\ref{fig:pitch_dist}(a), exhibits a peak at a value slightly larger than the body diameter, $d$,  and the width is approximately $d/2$. This large surface separation can only be understood by a significant wobbling dynamics, or a preferred alignment of the body toward or away from the wall, because in a perfectly parallel wall alignment state, the preferred distance should be $y_{cm} \approx d/2$. Moreover, the figure suggest a weak dependence on the starting orientation of the cell. The displayed differences result mainly from lack of sufficient statistics. 

Quantitatively, the strong variation of the body orientation is illustrated in Fig.~\ref{fig:pitch_dist}(b) for the two  snapshots displayed in  Figure~\ref{fig:snapshot_cell_orient}. The latter emphasize the large orientational variations due to the particular (random) arrangement of flagella.  In case of the larger variations of $\theta_b$, the distribution function exhibits two peaks indicating a preferred orientation, which in three-dimension corresponds to the swimming direction precessing on a cone. However, this does not affect the overall orientation of the cell, since the average inclination angle is centered  around zero with little difference between the two realizations. The distribution function of the  inclination angle is presented in Fig.~\ref{fig:pitch_dist}(c). Despite the broad distribution of inclination angles with negative values  due to wobbling, the overall distribution function depends very little on the initial angle $\theta_s$, as it should be expected, and exhibits a maximum at a positive value. Again, deviations between data for different initial values are of statistical origin. Hence, our simulations reveal a preferred cell orientation toward the wall, in agreement with experimental results \cite{PhysRevXDiLeo,bian:19} and simulations,\cite{eise:16.1} but in contrast to theoretical calculations for a straight, no-wobbling cell,  \cite{Shum1725,pimponi_chinappi_gualtieri_casciola_2016} which predict an orientation away from the wall.

Interestingly, the large variations in body orientation have little effect on the orientation of the flagella bundle, and hence the overall cell orientation. Evidently, the bundle is quite stable and its rotation dictates the cell-body dynamics rather than vice versa. 

\begin{figure}[t]
	\includegraphics[width=\columnwidth]{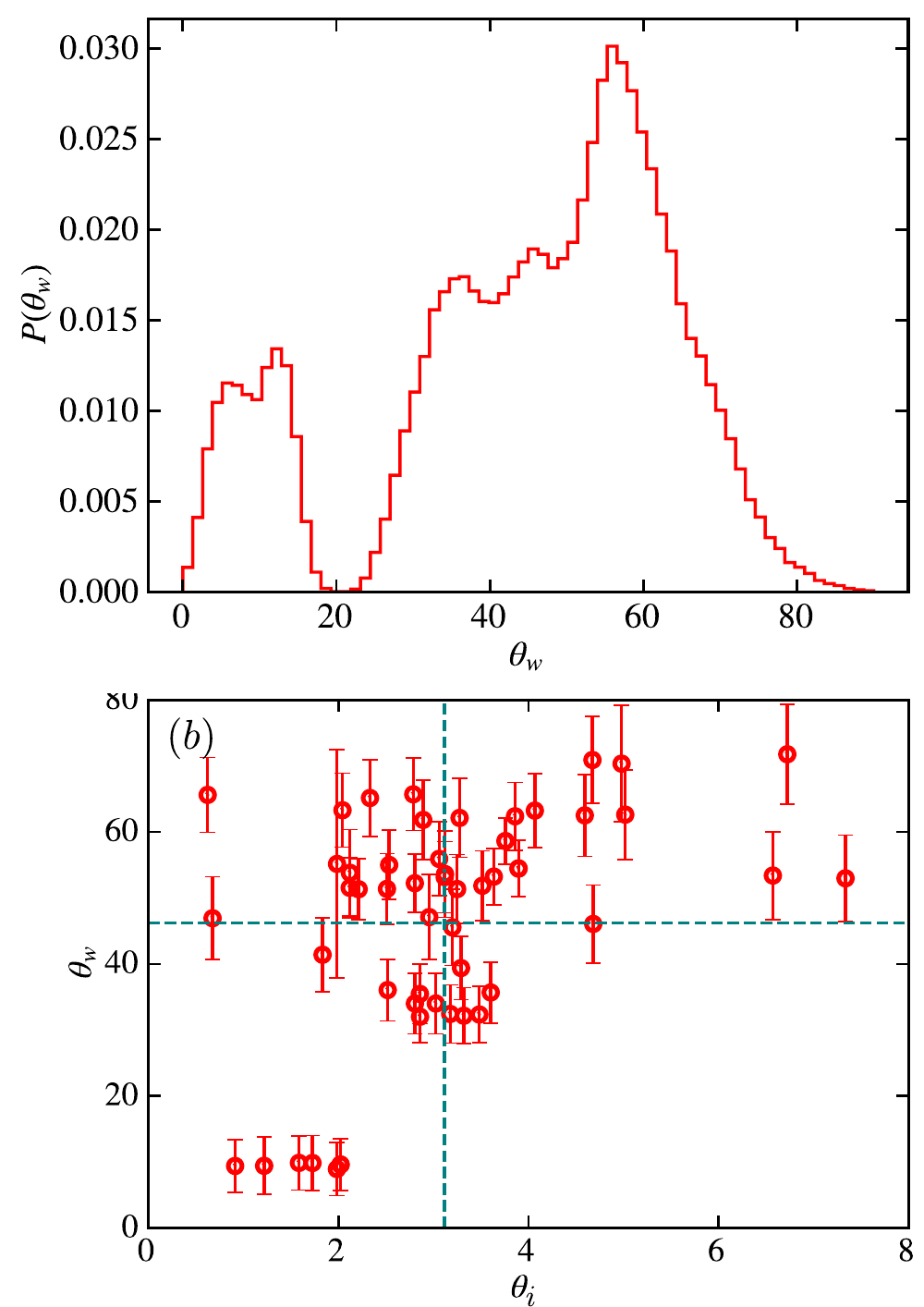}
	\caption{(a) Distribution function of the wobbling angle $\theta_w$ for the applied realizations. (b) Scatter plot of the wobbling  and inclination angle $\theta_i$ at the wall  (different initial angles and their  various realizations). The dashed lines indicate the average values.}
	\label{fig:wobb_pitch}
\end{figure}

To quantify the wobbling motion, we introduce the wobbling angle $\theta_w$ as the angle between the major axis of the body and the cell's swimming direction (main axis of the cell's inertia tensor). Figure~\ref{fig:wobb_pitch}(a) displays the distribution function $P(\theta_w)$ of the wobbling angle for the various realizations.  Evidently, we obtain a very broad distribution with the mean value  $\theta_{w} = \ang{46.2} \pm \ang{17.8}$. Experiments yield the smaller average angle $\theta_w  \approx 30^{\circ}$, although the range of angles is comparable.\cite{bian:19} Similarly, simulations of long and highly flagellated cells yield an average value of approximately $30^{\circ}$.\cite{eise:16.1} Interestingly, we find a roughly bimodal distribution with a peak  at about $10^{\circ}$ and a very broad peak at about $55^{\circ}$, separated by  a pronounced gap  at $20^{\circ}$. This could be related  to the preferentially organize of the bundle, which is rather well aligned with the cell body, or rather oblique to it. This is in contrast to  experimental distributions, which  show a high probability for lower angles $<20^{\circ}$ and no gap.  The discrepancy may be a consequence of the chosen fixed number of flagella, whereas {\em E. coli} planktonic cells possess approximately $4-7$ flagella. Simulations of swarmer-type cells, with a larger number of flagella, show that the number of flagella  matters for bundle formation, and no gap or high probability for small $\theta_w$ has been found.\cite{eise:16.1}   Thus, the role of the number of flagella in bundle formation and orientation, from a few to many, needs more detailed investigations.

The scatter plot in Fig.~\ref{fig:wobb_pitch}(b)
collects mean values of the inclination and wobbling angle for the various realizations.  For the mean value over all realizations we find  $\theta_{i} = \ang{3.1} \pm \ang{1.4}$. As pointed out before, the positive inclination angle is in agreement with experiments,\cite{PhysRevXDiLeo,bian:19} although our value is somewhat smaller.  Simulation studies with a single flagellum aligned with the cell body yield the inclination angle $\theta_i \approx 10^{\circ}$.\cite{math:19} The larger value could be a consequence of the particular geometry of the applied {\em E. coli} model, where the helix radius increases  toward the rear end of the cell. Since the helix radius is larger than that of the cell, steric flagellum-surface interactions prevent parallel alignment with respect to the surface and imply an orientation toward the wall. 

\section{Summary and Conclusions} \label{sec:summary}

We have analyzed the entrapment dynamics of an {\em E. coli}-type cell at a no-slip wall by mesoscale hydrodynamic simulations. The random arrangement of (five) flagella on the cell body leads to a strong wobbling motion. The entrapment dynamics can be classified into three phases, a nearly straight approach, reorientation, and surface swimming.  
We have shown that swimmer-wall hydrodynamic  interactions hardly affect the cell orientation as it approaches the wall, since the initial and collisional orientations with respect to the surface change only weakly and can be attributed to fluctuations. The mechanism for the slow-down of the swimming velocity in the vicinity of the wall is less evident due to large scatter in our simulation data. Based on the theoretical models, we can neither rule out the frictional force by an approaching Stokeslet nor the force by the cell's force dipole, because both approaches yield quantitatively reasonable agreement with the simulation results.  
Similarly, the reorientation of the cell upon touching the wall can be attributed to hydrodynamics and/or a torque emerging  from the propulsion force.   The extracted hydrodynamic rotation frequency  is approximately three times larger than that  obtained in simulations. Similarly, rotation by  steric interactions is in good quantitative  agreement with simulation results. Due to uncertainties in the parameters of the applied theoretical approaches, no definite conclusion about a dominant mechanism is possible based on the simulation data, and both, hydrodynamic and non-hydrodynamic effects, could contribute simultaneously rather than a single mechanism only. A qualitative and quantitative understanding requires a detailed calculation with a  more adequate model of a flagellated cell in the vicinity of a wall.
Finally, cells swim smoothly along the wall on clockwise circular trajectories, with the major axis of the cell preferentially directed toward the wall.  This final state is independent of the initial condition. 

The cells exhibit a pronounced wobbling motion, both in bulk and adjacent to a wall. The magnitude of the wobbling angle is determined by the arrangement of the flagella on the cell surface, the further they are away from the poles of the cylindrical body, the stronger the wobbling. In general, the orientation of the flagella bundle dictates the swimming direction, and its rotation determines the alignment of the cell body rather than the other way around. However, the wobbling frequency is equal to the rotation frequency of the body, which is on average  $1/5$ of the bundle rotation frequency. 


The flagella bundle, which is in our case thinner than the body diameter, is pointing toward the wall, independent of the extent of cell-body angular variations. The cell body orientation is strongly affected by wobbling, but the overall cell orientation is toward the wall. This contradicts theoretical results, which suggest that the inclination of a  straight bacterium is  pointing away from the wall (negative inclination angle).\cite{Shum1725,pimponi_chinappi_gualtieri_casciola_2016} Our simulations suggest that the inclination angle is only weakly  affected by the wobbling angle, despite a broad distribution of $\theta_w$, the range of $\theta_i$ is rather small.  
Quantitatively, our simulations yield a somewhat smaller inclination angle than experiments using {\em E. Coli} bacteria.\cite{PhysRevXDiLeo,bian:19} The origin could be specific bacteria-wall interactions not captured in our model, such as particular adhesive sites on the bacteria surface, or interactions with unbundled short flagella. 

 A characteristics of flagellated bacteria is a large variation of their behavior, depending on various aspects such as number of flagella, flagella arrangement on the cell surface, formation of the bundle, etc.\cite{eise:16.1} Hence, simulation of a suitable ensemble and extraction of meaning full averages is rather demanding, specifically with a detailed bacterium model as employed in this study. Here, many more in-depth studies are required to achieve a quantitative understanding of swimming bacteria in bulk and at walls.    

In summary, our simulations provide insight into the scattering of flagellated bacteria at walls, in particular, into the relevance of  swimmer-wall  hydrodynamic interactions. We hope that our results, specifically the importance of wobbling, will stimulate comparable experimental studies.

\section*{Acknowledgements}

Financial support by the Deutsche Forschungsgemeinschaft (DFG) within the priority program SPP 1726 ``Microswimmers--from Single Particle Motion to Collective Behaviour'' is gratefully acknowledged.


\begin{mcitethebibliography}{70}
\providecommand*{\natexlab}[1]{#1}
\providecommand*{\mciteSetBstSublistMode}[1]{}
\providecommand*{\mciteSetBstMaxWidthForm}[2]{}
\providecommand*{\mciteBstWouldAddEndPuncttrue}
  {\def\EndOfBibitem{\unskip.}}
\providecommand*{\mciteBstWouldAddEndPunctfalse}
  {\let\EndOfBibitem\relax}
\providecommand*{\mciteSetBstMidEndSepPunct}[3]{}
\providecommand*{\mciteSetBstSublistLabelBeginEnd}[3]{}
\providecommand*{\EndOfBibitem}{}
\mciteSetBstSublistMode{f}
\mciteSetBstMaxWidthForm{subitem}
{(\emph{\alph{mcitesubitemcount}})}
\mciteSetBstSublistLabelBeginEnd{\mcitemaxwidthsubitemform\space}
{\relax}{\relax}

\bibitem[Costerton \emph{et~al.}(1995)Costerton, Lewandowski, Caldwell, Korber,
  and Lappin-Scott]{cost:95}
J.~W. Costerton, Z.~Lewandowski, D.~E. Caldwell, D.~R. Korber and H.~M.
  Lappin-Scott, \emph{Annu. Rev. Microbiol.}, 1995, \textbf{49}, 711\relax
\mciteBstWouldAddEndPuncttrue
\mciteSetBstMidEndSepPunct{\mcitedefaultmidpunct}
{\mcitedefaultendpunct}{\mcitedefaultseppunct}\relax
\EndOfBibitem
\bibitem[Lauga(2016)]{laug:16}
E.~Lauga, \emph{Annu. Rev. Fluid Mech.}, 2016, \textbf{48}, 105\relax
\mciteBstWouldAddEndPuncttrue
\mciteSetBstMidEndSepPunct{\mcitedefaultmidpunct}
{\mcitedefaultendpunct}{\mcitedefaultseppunct}\relax
\EndOfBibitem
\bibitem[Hartmann \emph{et~al.}(2019)Hartmann, Singh, Pearce, Mok, Song,
  D{\'\i}az-Pascual, Dunkel, and Drescher]{hart:19}
R.~Hartmann, P.~K. Singh, P.~Pearce, R.~Mok, B.~Song, F.~D{\'\i}az-Pascual,
  J.~Dunkel and K.~Drescher, \emph{Nat. Phys.}, 2019, \textbf{15}, 251\relax
\mciteBstWouldAddEndPuncttrue
\mciteSetBstMidEndSepPunct{\mcitedefaultmidpunct}
{\mcitedefaultendpunct}{\mcitedefaultseppunct}\relax
\EndOfBibitem
\bibitem[Flemming \emph{et~al.}(2016)Flemming, Wingender, Szewzyk, Steinberg,
  Rice, and Kjelleberg]{flem:16}
H.-C. Flemming, J.~Wingender, U.~Szewzyk, P.~Steinberg, S.~A. Rice and
  S.~Kjelleberg, \emph{Nat. Rev. Microbiol.}, 2016, \textbf{14}, 563 EP\relax
\mciteBstWouldAddEndPuncttrue
\mciteSetBstMidEndSepPunct{\mcitedefaultmidpunct}
{\mcitedefaultendpunct}{\mcitedefaultseppunct}\relax
\EndOfBibitem
\bibitem[Koo \emph{et~al.}(2017)Koo, Allan, Howlin, Stoodley, and
  Hall-Stoodley]{koo:17}
H.~Koo, R.~N. Allan, R.~P. Howlin, P.~Stoodley and L.~Hall-Stoodley, \emph{Nat.
  Rev. Microbiol.}, 2017, \textbf{15}, 740 EP\relax
\mciteBstWouldAddEndPuncttrue
\mciteSetBstMidEndSepPunct{\mcitedefaultmidpunct}
{\mcitedefaultendpunct}{\mcitedefaultseppunct}\relax
\EndOfBibitem
\bibitem[Berke \emph{et~al.}(2008)Berke, Turner, Berg, and
  Lauga]{PhysRevLett.101.038102}
A.~P. Berke, L.~Turner, H.~C. Berg and E.~Lauga, \emph{Phys. Rev. Lett.}, 2008,
  \textbf{101}, 038102\relax
\mciteBstWouldAddEndPuncttrue
\mciteSetBstMidEndSepPunct{\mcitedefaultmidpunct}
{\mcitedefaultendpunct}{\mcitedefaultseppunct}\relax
\EndOfBibitem
\bibitem[Rothschild(1963)]{Rothschild1963}
Rothschild, \emph{Nature}, 1963, \textbf{198}, 1221\relax
\mciteBstWouldAddEndPuncttrue
\mciteSetBstMidEndSepPunct{\mcitedefaultmidpunct}
{\mcitedefaultendpunct}{\mcitedefaultseppunct}\relax
\EndOfBibitem
\bibitem[Spagnolie and Lauga(2012)]{spagnolie_lauga_2012}
S.~E. Spagnolie and E.~Lauga, \emph{J. Fluid Mech.}, 2012, \textbf{700},
  105–147\relax
\mciteBstWouldAddEndPuncttrue
\mciteSetBstMidEndSepPunct{\mcitedefaultmidpunct}
{\mcitedefaultendpunct}{\mcitedefaultseppunct}\relax
\EndOfBibitem
\bibitem[Elgeti and Gompper(2016)]{elge:16}
J.~Elgeti and G.~Gompper, \emph{Eur. Phys. J. Spec. Top.}, 2016, \textbf{225},
  2333\relax
\mciteBstWouldAddEndPuncttrue
\mciteSetBstMidEndSepPunct{\mcitedefaultmidpunct}
{\mcitedefaultendpunct}{\mcitedefaultseppunct}\relax
\EndOfBibitem
\bibitem[Elgeti \emph{et~al.}(2015)Elgeti, Winkler, and Gompper]{elge:15}
J.~Elgeti, R.~G. Winkler and G.~Gompper, \emph{Rep. Prog. Phys.}, 2015,
  \textbf{78}, 056601\relax
\mciteBstWouldAddEndPuncttrue
\mciteSetBstMidEndSepPunct{\mcitedefaultmidpunct}
{\mcitedefaultendpunct}{\mcitedefaultseppunct}\relax
\EndOfBibitem
\bibitem[Sipos \emph{et~al.}(2015)Sipos, Nagy, Di~Leonardo, and
  Galajda]{PhysRevLett.114.258104}
O.~Sipos, K.~Nagy, R.~Di~Leonardo and P.~Galajda, \emph{Phys. Rev. Lett.},
  2015, \textbf{114}, 258104\relax
\mciteBstWouldAddEndPuncttrue
\mciteSetBstMidEndSepPunct{\mcitedefaultmidpunct}
{\mcitedefaultendpunct}{\mcitedefaultseppunct}\relax
\EndOfBibitem
\bibitem[Happel and Brenner(1983)]{Happel1983}
J.~Happel and H.~Brenner, \emph{Low {Reynolds} number hydrodynamics: with
  special applications to particulate media}, Springer Science \& Business
  Media, 1983\relax
\mciteBstWouldAddEndPuncttrue
\mciteSetBstMidEndSepPunct{\mcitedefaultmidpunct}
{\mcitedefaultendpunct}{\mcitedefaultseppunct}\relax
\EndOfBibitem
\bibitem[Das and Lauga(2019)]{das:19.1}
D.~Das and E.~Lauga, \emph{Phys. Rev. E}, 2019, \textbf{100}, 043117\relax
\mciteBstWouldAddEndPuncttrue
\mciteSetBstMidEndSepPunct{\mcitedefaultmidpunct}
{\mcitedefaultendpunct}{\mcitedefaultseppunct}\relax
\EndOfBibitem
\bibitem[Lauga \emph{et~al.}(2006)Lauga, DiLuzio, Whitesides, and
  Stone]{BiophysJLauga}
E.~Lauga, W.~R. DiLuzio, G.~M. Whitesides and H.~A. Stone, \emph{Biophys. J.},
  2006, \textbf{90}, 400 -- 412\relax
\mciteBstWouldAddEndPuncttrue
\mciteSetBstMidEndSepPunct{\mcitedefaultmidpunct}
{\mcitedefaultendpunct}{\mcitedefaultseppunct}\relax
\EndOfBibitem
\bibitem[Li and Tang(2009)]{PhysRevLett.103.078101}
G.~Li and J.~X. Tang, \emph{Phys. Rev. Lett.}, 2009, \textbf{103}, 078101\relax
\mciteBstWouldAddEndPuncttrue
\mciteSetBstMidEndSepPunct{\mcitedefaultmidpunct}
{\mcitedefaultendpunct}{\mcitedefaultseppunct}\relax
\EndOfBibitem
\bibitem[Li \emph{et~al.}(2011)Li, Bensson, Nisimova, Munger, Mahautmr, Tang,
  Maxey, and Brun]{PhysRevE.84.041932}
G.~Li, J.~Bensson, L.~Nisimova, D.~Munger, P.~Mahautmr, J.~X. Tang, M.~R. Maxey
  and Y.~V. Brun, \emph{Phys. Rev. E}, 2011, \textbf{84}, 041932\relax
\mciteBstWouldAddEndPuncttrue
\mciteSetBstMidEndSepPunct{\mcitedefaultmidpunct}
{\mcitedefaultendpunct}{\mcitedefaultseppunct}\relax
\EndOfBibitem
\bibitem[Elgeti and Gompper(2009)]{elge:09}
J.~Elgeti and G.~Gompper, \emph{EPL}, 2009, \textbf{85}, 38002\relax
\mciteBstWouldAddEndPuncttrue
\mciteSetBstMidEndSepPunct{\mcitedefaultmidpunct}
{\mcitedefaultendpunct}{\mcitedefaultseppunct}\relax
\EndOfBibitem
\bibitem[Shum \emph{et~al.}(2010)Shum, Gaffney, and Smith]{Shum1725}
H.~Shum, E.~A. Gaffney and D.~J. Smith, \emph{Proc. R. Soc. A}, 2010,
  \textbf{466}, 1725--1748\relax
\mciteBstWouldAddEndPuncttrue
\mciteSetBstMidEndSepPunct{\mcitedefaultmidpunct}
{\mcitedefaultendpunct}{\mcitedefaultseppunct}\relax
\EndOfBibitem
\bibitem[Pimponi \emph{et~al.}(2016)Pimponi, Chinappi, Gualtieri, and
  Casciola]{pimponi_chinappi_gualtieri_casciola_2016}
D.~Pimponi, M.~Chinappi, P.~Gualtieri and C.~M. Casciola, \emph{J. Fluid
  Mech.}, 2016, \textbf{789}, 514–533\relax
\mciteBstWouldAddEndPuncttrue
\mciteSetBstMidEndSepPunct{\mcitedefaultmidpunct}
{\mcitedefaultendpunct}{\mcitedefaultseppunct}\relax
\EndOfBibitem
\bibitem[Hu \emph{et~al.}(2015)Hu, Wysocki, Winkler, and Gompper]{HuSciRep2015}
J.~Hu, A.~Wysocki, R.~G. Winkler and G.~Gompper, \emph{Sci. Rep.}, 2015,
  \textbf{5}, 9586\relax
\mciteBstWouldAddEndPuncttrue
\mciteSetBstMidEndSepPunct{\mcitedefaultmidpunct}
{\mcitedefaultendpunct}{\mcitedefaultseppunct}\relax
\EndOfBibitem
\bibitem[Eisenstecken \emph{et~al.}(2016)Eisenstecken, Hu, and
  Winkler]{eise:16.1}
T.~Eisenstecken, J.~Hu and R.~G. Winkler, \emph{Soft Matter}, 2016,
  \textbf{12}, 8316\relax
\mciteBstWouldAddEndPuncttrue
\mciteSetBstMidEndSepPunct{\mcitedefaultmidpunct}
{\mcitedefaultendpunct}{\mcitedefaultseppunct}\relax
\EndOfBibitem
\bibitem[Bianchi \emph{et~al.}(2017)Bianchi, Saglimbeni, and
  Di~Leonardo]{PhysRevXDiLeo}
S.~Bianchi, F.~Saglimbeni and R.~Di~Leonardo, \emph{Phys. Rev. X}, 2017,
  \textbf{7}, 011010\relax
\mciteBstWouldAddEndPuncttrue
\mciteSetBstMidEndSepPunct{\mcitedefaultmidpunct}
{\mcitedefaultendpunct}{\mcitedefaultseppunct}\relax
\EndOfBibitem
\bibitem[Bianchi \emph{et~al.}(2019)Bianchi, Saglimbeni, Frangipane,
  Dell'Arciprete, and Di~Leonardo]{bian:19}
S.~Bianchi, F.~Saglimbeni, G.~Frangipane, D.~Dell'Arciprete and R.~Di~Leonardo,
  \emph{Soft Matter}, 2019, \textbf{15}, 3397\relax
\mciteBstWouldAddEndPuncttrue
\mciteSetBstMidEndSepPunct{\mcitedefaultmidpunct}
{\mcitedefaultendpunct}{\mcitedefaultseppunct}\relax
\EndOfBibitem
\bibitem[Palagi and Fischer(2018)]{pala:18}
S.~Palagi and P.~Fischer, \emph{Nat. Rev. Mater.}, 2018, \textbf{3}, 113\relax
\mciteBstWouldAddEndPuncttrue
\mciteSetBstMidEndSepPunct{\mcitedefaultmidpunct}
{\mcitedefaultendpunct}{\mcitedefaultseppunct}\relax
\EndOfBibitem
\bibitem[Gompper \emph{et~al.}(2020)Gompper, Winkler, Speck, Solon, Nardini,
  Peruani, L{\"o}wen, Golestanian, Kaupp, Alvarez, Ki{\o}rboe, Lauga, Poon,
  DeSimone, Mui{\~n}os-Landin, Fischer, S{\"o}ker, Cichos, Kapral, Gaspard,
  Ripoll, Sagues, Doostmohammadi, Yeomans, Aranson, Bechinger, Stark,
  Hemelrijk, Nedelec, Sarkar, Aryaksama, Lacroix, Duclos, Yashunsky, Silberzan,
  Arroyo, and Kale]{gomp:20}
G.~Gompper, R.~G. Winkler, T.~Speck, A.~Solon, C.~Nardini, F.~Peruani,
  H.~L{\"o}wen, R.~Golestanian, U.~B. Kaupp, L.~Alvarez, T.~Ki{\o}rboe,
  E.~Lauga, W.~C.~K. Poon, A.~DeSimone, S.~Mui{\~n}os-Landin, A.~Fischer, N.~A.
  S{\"o}ker, F.~Cichos, R.~Kapral, P.~Gaspard, M.~Ripoll, F.~Sagues,
  A.~Doostmohammadi, J.~M. Yeomans, I.~S. Aranson, C.~Bechinger, H.~Stark,
  C.~K. Hemelrijk, F.~J. Nedelec, T.~Sarkar, T.~Aryaksama, M.~Lacroix,
  G.~Duclos, V.~Yashunsky, P.~Silberzan, M.~Arroyo and S.~Kale, \emph{J. Phys:
  Condens. Matter}, 2020, \textbf{32}, 193001\relax
\mciteBstWouldAddEndPuncttrue
\mciteSetBstMidEndSepPunct{\mcitedefaultmidpunct}
{\mcitedefaultendpunct}{\mcitedefaultseppunct}\relax
\EndOfBibitem
\bibitem[Hu \emph{et~al.}(2015)Hu, Yang, Gompper, and Winkler]{hu:15.1}
J.~Hu, M.~Yang, G.~Gompper and R.~G. Winkler, \emph{Soft Matter}, 2015,
  \textbf{11}, 7867\relax
\mciteBstWouldAddEndPuncttrue
\mciteSetBstMidEndSepPunct{\mcitedefaultmidpunct}
{\mcitedefaultendpunct}{\mcitedefaultseppunct}\relax
\EndOfBibitem
\bibitem[Ripoll \emph{et~al.}(2004)Ripoll, Mussawisade, Winkler, and
  Gompper]{ripo:04}
M.~Ripoll, K.~Mussawisade, R.~G. Winkler and G.~Gompper, \emph{Europhys.
  Lett.}, 2004, \textbf{68}, 106\relax
\mciteBstWouldAddEndPuncttrue
\mciteSetBstMidEndSepPunct{\mcitedefaultmidpunct}
{\mcitedefaultendpunct}{\mcitedefaultseppunct}\relax
\EndOfBibitem
\bibitem[Malevanets and Kapral(1999)]{male:99}
A.~Malevanets and R.~Kapral, \emph{J. Chem. Phys.}, 1999, \textbf{110},
  8605\relax
\mciteBstWouldAddEndPuncttrue
\mciteSetBstMidEndSepPunct{\mcitedefaultmidpunct}
{\mcitedefaultendpunct}{\mcitedefaultseppunct}\relax
\EndOfBibitem
\bibitem[Kapral(2008)]{kapr:08}
R.~Kapral, \emph{Adv. Chem. Phys.}, 2008, \textbf{140}, 89\relax
\mciteBstWouldAddEndPuncttrue
\mciteSetBstMidEndSepPunct{\mcitedefaultmidpunct}
{\mcitedefaultendpunct}{\mcitedefaultseppunct}\relax
\EndOfBibitem
\bibitem[Gompper \emph{et~al.}(2009)Gompper, Ihle, Kroll, and Winkler]{gomp:09}
G.~Gompper, T.~Ihle, D.~M. Kroll and R.~G. Winkler, \emph{Adv. Polym. Sci.},
  2009, \textbf{221}, 1\relax
\mciteBstWouldAddEndPuncttrue
\mciteSetBstMidEndSepPunct{\mcitedefaultmidpunct}
{\mcitedefaultendpunct}{\mcitedefaultseppunct}\relax
\EndOfBibitem
\bibitem[G{\"o}tze and Gompper(2010)]{goet:10}
I.~O. G{\"o}tze and G.~Gompper, \emph{Phys. Rev. E}, 2010, \textbf{82},
  041921\relax
\mciteBstWouldAddEndPuncttrue
\mciteSetBstMidEndSepPunct{\mcitedefaultmidpunct}
{\mcitedefaultendpunct}{\mcitedefaultseppunct}\relax
\EndOfBibitem
\bibitem[Earl \emph{et~al.}(2007)Earl, Pooley, Ryder, Bredberg, and
  Yeomans]{earl:07}
D.~J. Earl, C.~M. Pooley, J.~F. Ryder, I.~Bredberg and J.~M. Yeomans, \emph{J.
  Chem. Phys.}, 2007, \textbf{126}, 064703\relax
\mciteBstWouldAddEndPuncttrue
\mciteSetBstMidEndSepPunct{\mcitedefaultmidpunct}
{\mcitedefaultendpunct}{\mcitedefaultseppunct}\relax
\EndOfBibitem
\bibitem[Elgeti \emph{et~al.}(2010)Elgeti, Kaupp, and Gompper]{elge:10}
J.~Elgeti, U.~B. Kaupp and G.~Gompper, \emph{Biophys. J.}, 2010, \textbf{99},
  1018\relax
\mciteBstWouldAddEndPuncttrue
\mciteSetBstMidEndSepPunct{\mcitedefaultmidpunct}
{\mcitedefaultendpunct}{\mcitedefaultseppunct}\relax
\EndOfBibitem
\bibitem[Babu and Stark(2012)]{babu:12}
S.~B. Babu and H.~Stark, \emph{New J. Phys.}, 2012, \textbf{14}, 085012\relax
\mciteBstWouldAddEndPuncttrue
\mciteSetBstMidEndSepPunct{\mcitedefaultmidpunct}
{\mcitedefaultendpunct}{\mcitedefaultseppunct}\relax
\EndOfBibitem
\bibitem[Elgeti and Gompper(2013)]{elge:13}
J.~Elgeti and G.~Gompper, \emph{Proc. Natl. Acad. Sci. USA}, 2013,
  \textbf{110}, 4470\relax
\mciteBstWouldAddEndPuncttrue
\mciteSetBstMidEndSepPunct{\mcitedefaultmidpunct}
{\mcitedefaultendpunct}{\mcitedefaultseppunct}\relax
\EndOfBibitem
\bibitem[Theers and Winkler(2013)]{thee:13}
M.~Theers and R.~G. Winkler, \emph{Phys. Rev. E}, 2013, \textbf{88},
  023012\relax
\mciteBstWouldAddEndPuncttrue
\mciteSetBstMidEndSepPunct{\mcitedefaultmidpunct}
{\mcitedefaultendpunct}{\mcitedefaultseppunct}\relax
\EndOfBibitem
\bibitem[Yang and Ripoll(2014)]{yang:14}
M.~Yang and M.~Ripoll, \emph{Soft Matter}, 2014, \textbf{10}, 1006\relax
\mciteBstWouldAddEndPuncttrue
\mciteSetBstMidEndSepPunct{\mcitedefaultmidpunct}
{\mcitedefaultendpunct}{\mcitedefaultseppunct}\relax
\EndOfBibitem
\bibitem[Z{\"o}ttl and Stark(2014)]{zoet:14}
A.~Z{\"o}ttl and H.~Stark, \emph{Phys. Rev. Lett.}, 2014, \textbf{112},
  118101\relax
\mciteBstWouldAddEndPuncttrue
\mciteSetBstMidEndSepPunct{\mcitedefaultmidpunct}
{\mcitedefaultendpunct}{\mcitedefaultseppunct}\relax
\EndOfBibitem
\bibitem[Qi \emph{et~al.}(2020)Qi, Westphal, Gompper, and Winkler]{qi:20}
K.~Qi, E.~Westphal, G.~Gompper and R.~G. Winkler, \emph{Phys. Rev. Lett.},
  2020, \textbf{124}, 068001\relax
\mciteBstWouldAddEndPuncttrue
\mciteSetBstMidEndSepPunct{\mcitedefaultmidpunct}
{\mcitedefaultendpunct}{\mcitedefaultseppunct}\relax
\EndOfBibitem
\bibitem[Yang \emph{et~al.}(2008)Yang, Elgeti, and Gompper]{yang:08}
Y.~Yang, J.~Elgeti and G.~Gompper, \emph{Phys. Rev. E}, 2008, \textbf{78},
  061903\relax
\mciteBstWouldAddEndPuncttrue
\mciteSetBstMidEndSepPunct{\mcitedefaultmidpunct}
{\mcitedefaultendpunct}{\mcitedefaultseppunct}\relax
\EndOfBibitem
\bibitem[Reigh \emph{et~al.}(2012)Reigh, Winkler, and Gompper]{reig:12}
S.~Y. Reigh, R.~G. Winkler and G.~Gompper, \emph{Soft Matter}, 2012,
  \textbf{8}, 4363\relax
\mciteBstWouldAddEndPuncttrue
\mciteSetBstMidEndSepPunct{\mcitedefaultmidpunct}
{\mcitedefaultendpunct}{\mcitedefaultseppunct}\relax
\EndOfBibitem
\bibitem[Reigh \emph{et~al.}(2013)Reigh, Winkler, and Gompper]{reig:13}
S.~Y. Reigh, R.~G. Winkler and G.~Gompper, \emph{{PLoS} {ONE}}, 2013,
  \textbf{8}, e70868\relax
\mciteBstWouldAddEndPuncttrue
\mciteSetBstMidEndSepPunct{\mcitedefaultmidpunct}
{\mcitedefaultendpunct}{\mcitedefaultseppunct}\relax
\EndOfBibitem
\bibitem[Theers \emph{et~al.}(2018)Theers, Westphal, Qi, Winkler, and
  Gompper]{thee:18}
M.~Theers, E.~Westphal, K.~Qi, R.~G. Winkler and G.~Gompper, \emph{Soft
  Matter}, 2018, \textbf{14}, 8590\relax
\mciteBstWouldAddEndPuncttrue
\mciteSetBstMidEndSepPunct{\mcitedefaultmidpunct}
{\mcitedefaultendpunct}{\mcitedefaultseppunct}\relax
\EndOfBibitem
\bibitem[Drescher \emph{et~al.}(2011)Drescher, Dunkel, Cisneros, Ganguly, and
  Goldstein]{dres:11}
K.~Drescher, J.~Dunkel, L.~H. Cisneros, S.~Ganguly and R.~E. Goldstein,
  \emph{Proc. Natl. Acad. Sci. USA}, 2011, \textbf{108}, 10940\relax
\mciteBstWouldAddEndPuncttrue
\mciteSetBstMidEndSepPunct{\mcitedefaultmidpunct}
{\mcitedefaultendpunct}{\mcitedefaultseppunct}\relax
\EndOfBibitem
\bibitem[Hyon \emph{et~al.}(2012)Hyon, Marcos, Powers, Stocker, and
  Fu]{hyon:12}
Y.~Hyon, u.~Marcos, T.~R. Powers, R.~Stocker and H.~C. Fu, \emph{J. Fluid
  Mech.}, 2012, \textbf{705}, 58--76\relax
\mciteBstWouldAddEndPuncttrue
\mciteSetBstMidEndSepPunct{\mcitedefaultmidpunct}
{\mcitedefaultendpunct}{\mcitedefaultseppunct}\relax
\EndOfBibitem
\bibitem[Yamakawa and Yoshizaki(1997)]{HWCh1997}
\emph{Helical Wormlike Chains in Polymer Solutions}, ed. H.~Yamakawa and
  T.~Yoshizaki, Springer Verlag, Berlin Heidelberg, 1997\relax
\mciteBstWouldAddEndPuncttrue
\mciteSetBstMidEndSepPunct{\mcitedefaultmidpunct}
{\mcitedefaultendpunct}{\mcitedefaultseppunct}\relax
\EndOfBibitem
\bibitem[Vogel and Stark(2010)]{Vogel2010}
R.~Vogel and H.~Stark, \emph{Eur. Phys. J. E}, 2010, \textbf{33},
  259--271\relax
\mciteBstWouldAddEndPuncttrue
\mciteSetBstMidEndSepPunct{\mcitedefaultmidpunct}
{\mcitedefaultendpunct}{\mcitedefaultseppunct}\relax
\EndOfBibitem
\bibitem[Darnton \emph{et~al.}(2007)Darnton, Turner, Rojevsky, and
  Berg]{Darnton1756}
N.~C. Darnton, L.~Turner, S.~Rojevsky and H.~C. Berg, \emph{J. Bacteriol.},
  2007, \textbf{189}, 1756--1764\relax
\mciteBstWouldAddEndPuncttrue
\mciteSetBstMidEndSepPunct{\mcitedefaultmidpunct}
{\mcitedefaultendpunct}{\mcitedefaultseppunct}\relax
\EndOfBibitem
\bibitem[Allen and Tildesley(1987)]{alle:87}
M.~P. Allen and D.~J. Tildesley, \emph{Computer Simulation of Liquids},
  Clarendon Press, Oxford, 1987\relax
\mciteBstWouldAddEndPuncttrue
\mciteSetBstMidEndSepPunct{\mcitedefaultmidpunct}
{\mcitedefaultendpunct}{\mcitedefaultseppunct}\relax
\EndOfBibitem
\bibitem[Theers \emph{et~al.}(2016)Theers, Westphal, Gompper, and
  Winkler]{thee:16}
M.~Theers, E.~Westphal, G.~Gompper and R.~G. Winkler, \emph{Phys. Rev. E},
  2016, \textbf{93}, 032604\relax
\mciteBstWouldAddEndPuncttrue
\mciteSetBstMidEndSepPunct{\mcitedefaultmidpunct}
{\mcitedefaultendpunct}{\mcitedefaultseppunct}\relax
\EndOfBibitem
\bibitem[Noguchi and Gompper(2008)]{nogu:08}
H.~Noguchi and G.~Gompper, \emph{Phys. Rev. E}, 2008, \textbf{78}, 016706\relax
\mciteBstWouldAddEndPuncttrue
\mciteSetBstMidEndSepPunct{\mcitedefaultmidpunct}
{\mcitedefaultendpunct}{\mcitedefaultseppunct}\relax
\EndOfBibitem
\bibitem[Theers and Winkler(2014)]{thee:14}
M.~Theers and R.~G. Winkler, \emph{Soft Matter}, 2014, \textbf{10}, 5894\relax
\mciteBstWouldAddEndPuncttrue
\mciteSetBstMidEndSepPunct{\mcitedefaultmidpunct}
{\mcitedefaultendpunct}{\mcitedefaultseppunct}\relax
\EndOfBibitem
\bibitem[Ihle and Kroll(2003)]{ihle:03}
T.~Ihle and D.~M. Kroll, \emph{Phys. Rev. E}, 2003, \textbf{67}, 066705\relax
\mciteBstWouldAddEndPuncttrue
\mciteSetBstMidEndSepPunct{\mcitedefaultmidpunct}
{\mcitedefaultendpunct}{\mcitedefaultseppunct}\relax
\EndOfBibitem
\bibitem[Huang \emph{et~al.}(2015)Huang, Varghese, Gompper, and
  Winkler]{huan:15}
C.-C. Huang, A.~Varghese, G.~Gompper and R.~G. Winkler, \emph{Phys. Rev. E},
  2015, \textbf{91}, 013310\relax
\mciteBstWouldAddEndPuncttrue
\mciteSetBstMidEndSepPunct{\mcitedefaultmidpunct}
{\mcitedefaultendpunct}{\mcitedefaultseppunct}\relax
\EndOfBibitem
\bibitem[Mussawisade \emph{et~al.}(2005)Mussawisade, Ripoll, Winkler, and
  Gompper]{muss:05}
K.~Mussawisade, M.~Ripoll, R.~G. Winkler and G.~Gompper, \emph{J. Chem. Phys.},
  2005, \textbf{123}, 144905\relax
\mciteBstWouldAddEndPuncttrue
\mciteSetBstMidEndSepPunct{\mcitedefaultmidpunct}
{\mcitedefaultendpunct}{\mcitedefaultseppunct}\relax
\EndOfBibitem
\bibitem[Poblete \emph{et~al.}(2014)Poblete, Wysocki, Gompper, and
  Winkler]{pobl:14}
S.~Poblete, A.~Wysocki, G.~Gompper and R.~G. Winkler, \emph{Phys. Rev. E},
  2014, \textbf{90}, 033314\relax
\mciteBstWouldAddEndPuncttrue
\mciteSetBstMidEndSepPunct{\mcitedefaultmidpunct}
{\mcitedefaultendpunct}{\mcitedefaultseppunct}\relax
\EndOfBibitem
\bibitem[Lamura \emph{et~al.}(2001)Lamura, Gompper, Ihle, and Kroll]{lamu:01}
A.~Lamura, G.~Gompper, T.~Ihle and D.~M. Kroll, \emph{Europhys. Lett.}, 2001,
  \textbf{56}, 319--325\relax
\mciteBstWouldAddEndPuncttrue
\mciteSetBstMidEndSepPunct{\mcitedefaultmidpunct}
{\mcitedefaultendpunct}{\mcitedefaultseppunct}\relax
\EndOfBibitem
\bibitem[Berry and Berg(1997)]{Berry14433}
R.~M. Berry and H.~C. Berg, \emph{Proc. Natl. Acad. Sci. USA}, 1997,
  \textbf{94}, 14433--14437\relax
\mciteBstWouldAddEndPuncttrue
\mciteSetBstMidEndSepPunct{\mcitedefaultmidpunct}
{\mcitedefaultendpunct}{\mcitedefaultseppunct}\relax
\EndOfBibitem
\bibitem[Das and Lauga(2018)]{das:18.2}
D.~Das and E.~Lauga, \emph{Soft Matter}, 2018, \textbf{14}, 5955\relax
\mciteBstWouldAddEndPuncttrue
\mciteSetBstMidEndSepPunct{\mcitedefaultmidpunct}
{\mcitedefaultendpunct}{\mcitedefaultseppunct}\relax
\EndOfBibitem
\bibitem[Liu \emph{et~al.}(2014)Liu, Gulino, Morse, Tang, Powers, and
  Breuer]{liu:14}
B.~Liu, M.~Gulino, M.~Morse, J.~X. Tang, T.~R. Powers and K.~S. Breuer,
  \emph{Proc. Natl. Acad. Sci. USA}, 2014, \textbf{111}, 11252\relax
\mciteBstWouldAddEndPuncttrue
\mciteSetBstMidEndSepPunct{\mcitedefaultmidpunct}
{\mcitedefaultendpunct}{\mcitedefaultseppunct}\relax
\EndOfBibitem
\bibitem[Darnton \emph{et~al.}(2007)Darnton, Turner, Rojevsky, and
  Berg]{darn:07.1}
N.~C. Darnton, L.~Turner, S.~Rojevsky and H.~C. Berg, \emph{J. Bacteriol.},
  2007, \textbf{189}, 1756--1764\relax
\mciteBstWouldAddEndPuncttrue
\mciteSetBstMidEndSepPunct{\mcitedefaultmidpunct}
{\mcitedefaultendpunct}{\mcitedefaultseppunct}\relax
\EndOfBibitem
\bibitem[Turner \emph{et~al.}(2010)Turner, Zhang, Darnton, and Berg]{turn:10}
L.~Turner, R.~Zhang, N.~C. Darnton and H.~C. Berg, \emph{J. Bacteriol.}, 2010,
  \textbf{192}, 3259\relax
\mciteBstWouldAddEndPuncttrue
\mciteSetBstMidEndSepPunct{\mcitedefaultmidpunct}
{\mcitedefaultendpunct}{\mcitedefaultseppunct}\relax
\EndOfBibitem
\bibitem[Patteson \emph{et~al.}(2015)Patteson, Gopinath, Goulian, and
  Arratia]{Patteson_2015}
A.~E. Patteson, A.~Gopinath, M.~Goulian and P.~E. Arratia, \emph{Sci. Rep.},
  2015, \textbf{5}, 15761\relax
\mciteBstWouldAddEndPuncttrue
\mciteSetBstMidEndSepPunct{\mcitedefaultmidpunct}
{\mcitedefaultendpunct}{\mcitedefaultseppunct}\relax
\EndOfBibitem
\bibitem[Hyon \emph{et~al.}(2012)Hyon, Marcos, Powers, Stocker, and
  Fu]{hyon_marcos_powers_stocker_fu_2012}
Y.~Hyon, Marcos, T.~R. Powers, R.~Stocker and H.~C. Fu, \emph{J. Fluid Mech.},
  2012, \textbf{705}, 58–76\relax
\mciteBstWouldAddEndPuncttrue
\mciteSetBstMidEndSepPunct{\mcitedefaultmidpunct}
{\mcitedefaultendpunct}{\mcitedefaultseppunct}\relax
\EndOfBibitem
\bibitem[Frymier \emph{et~al.}(1995)Frymier, Ford, Berg, and Cummings]{frym:95}
P.~D. Frymier, R.~M. Ford, H.~C. Berg and P.~T. Cummings, \emph{Proc. Natl.
  Acad. Sci. USA}, 1995, \textbf{92}, 6195\relax
\mciteBstWouldAddEndPuncttrue
\mciteSetBstMidEndSepPunct{\mcitedefaultmidpunct}
{\mcitedefaultendpunct}{\mcitedefaultseppunct}\relax
\EndOfBibitem
\bibitem[Ramia \emph{et~al.}(1993)Ramia, Tullock, and Phan-Thien]{rami:93}
M.~Ramia, D.~L. Tullock and N.~Phan-Thien, \emph{Biophys. J.}, 1993,
  \textbf{65}, 755\relax
\mciteBstWouldAddEndPuncttrue
\mciteSetBstMidEndSepPunct{\mcitedefaultmidpunct}
{\mcitedefaultendpunct}{\mcitedefaultseppunct}\relax
\EndOfBibitem
\bibitem[Kim and Karrila(1991)]{kim:91}
S.~Kim and S.~J. Karrila, \emph{Microhydrodynamics: principles and selected
  applications}, Butterworth-Heinemann, Boston, 1991\relax
\mciteBstWouldAddEndPuncttrue
\mciteSetBstMidEndSepPunct{\mcitedefaultmidpunct}
{\mcitedefaultendpunct}{\mcitedefaultseppunct}\relax
\EndOfBibitem
\bibitem[Winkler and Gompper(2018)]{wink:18}
R.~G. Winkler and G.~Gompper, \emph{Handbook of Materials Modeling: Methods:
  Theory and Modeling. Springer}, 2018,  1--20\relax
\mciteBstWouldAddEndPuncttrue
\mciteSetBstMidEndSepPunct{\mcitedefaultmidpunct}
{\mcitedefaultendpunct}{\mcitedefaultseppunct}\relax
\EndOfBibitem
\bibitem[Bianchi \emph{et~al.}(2017)Bianchi, Saglimbeni, and
  Di~Leonardo]{bian:17}
S.~Bianchi, F.~Saglimbeni and R.~Di~Leonardo, \emph{Phys. Rev. X}, 2017,
  \textbf{7}, 011010\relax
\mciteBstWouldAddEndPuncttrue
\mciteSetBstMidEndSepPunct{\mcitedefaultmidpunct}
{\mcitedefaultendpunct}{\mcitedefaultseppunct}\relax
\EndOfBibitem
\bibitem[Mathijssen \emph{et~al.}(2019)Mathijssen, Figueroa-Morales, Junot,
  Clement, Lindner, and Z{\"o}ttl]{math:19}
A.~J. T.~M. Mathijssen, N.~Figueroa-Morales, G.~Junot, E.~Clement, A.~Lindner
  and A.~Z{\"o}ttl, \emph{Nat. Commun.}, 2019, \textbf{20}, 3434\relax
\mciteBstWouldAddEndPuncttrue
\mciteSetBstMidEndSepPunct{\mcitedefaultmidpunct}
{\mcitedefaultendpunct}{\mcitedefaultseppunct}\relax
\EndOfBibitem
\end{mcitethebibliography}

\providecommand*{\mcitethebibliography}{\thebibliography}
\csname @ifundefined\endcsname{endmcitethebibliography}
{\let\endmcitethebibliography\endthebibliography}{}

\end{document}